\documentclass[aps,12pt]{article}
\usepackage{amsmath,amssymb,color,epsfig,cite}
\usepackage{xcolor}
\usepackage{mathrsfs}
\usepackage{authblk}

\usepackage{mathtools}

\usepackage{amsfonts}

\makeatletter
\@addtoreset{equation}{section}
\makeatother

\newcommand{\be}{\begin{equation}}
	\newcommand{\ee}{\end{equation}}
\newcommand{\bea} {\begin{eqnarray}}
		\newcommand{\eea}{\end{eqnarray}}
	
	\newcommand{\pb}[1]{\ensuremath{\partial_{#1}}}
	\newcommand{\cmmnt}[1]{}

	\def\0{{\sst{(0)}}}
	\def\1{{\sst{(1)}}}
	\def\2{{\sst{(2)}}}
	\def\3{{\sst{(3)}}}
	\def\4{{\sst{(4)}}}
	\def\5{{\sst{(5)}}}
	\def\6{{\sst{(6)}}}
	\def\7{{\sst{(7)}}}
	\def\8{{\sst{(8)}}}
	\def\sst#1{{\scriptscriptstyle #1}}

	\newcommand{\bx}{\ensuremath{\bar x}}

	\def\bx{{{\mathrm x}}}

	\def\vp{{\varphi}}

	\def\bx {{\bf x}}
	\newcommand{\bd}{\ensuremath{\bar{\Delta}}}

		\newcommand{\ben}{\begin{equation}}	
			\newcommand{\een}{\end{eqnarray}}	

	\def \p {\partial}
	\def\om{\omega}
	\def\bx{\bar{x}}
	\def\k{\kappa}
	\def\Om{\Omega}
	\def\ma{\mathfrak{a}}
	\def\mb{\mathfrak{b}}
	\def\mc{\mathfrak{c}}
	\def\G{\Gamma}
	\def\vk{\varkappa}

	\def\betap{\alpha_{+}}
	\def\betam{\alpha_{-}}

	\title{\Large{Tidal perturbations and Love Symmetry for five-dimensional  charged rotating black holes}}
	\author[1,2,3]{\small {M. Cveti\v{c}\thanks{e-mail: cvetic@physics.upenn.edu}}}  
	\author[4]{\small {M. A. Liao\thanks{e-mail: matheusalvesliao@gmail.com}}}
	\author[5]{\small {M. M. Stetsko\thanks{e-mail: mstetsko@gmail.com}}}
	\affil[1]{Department of Physics and Astronomy, University of Pennsylvania, Philadelphia, PA, 19104, USA}
	\affil[2]{Department of Mathematics, University of Pennsylvania, Philadelphia, PA, 19104, USA}
	\affil[3]{Center for Applied Mathematics and Theoretical Physics, University of Maribor, Maribor, Slovenia}
	\affil[4]{Departamento de Física, Universidade Federal da Paraíba, 58051-970 João Pessoa, PB, Brazil}
	\affil[5]{Department for Theoretical Physics, Ivan Franko National University of Lviv, Lviv, UA-79005, Ukraine}
	
\begin{document}
\vspace*{-2cm}
\begin{flushright}
	\texttt{UPR-1336-T}
\end{flushright}
{\let\newpage\relax\maketitle}
		\maketitle
		
		\begin{abstract}
			We investigate the tidal response of general five-dimensional (5D)  black holes of STU supergravity, which include as special cases important solutions such as the Myers-Perry, BMPV, 5D Reissner-Nordström, Kerr-Newman and dyonic black holes. Solutions are parameterized by their mass, two angular momenta and up to three U(1) charges. Love numbers and dissipation coefficients are obtained in the static and dynamic cases. In the latter scenario, we find new, nontrivial conditions, realized in important limiting cases of the theory, such as the BPS limit, where frequency-independent vanishing conditions are obtained. We also develop a ladder formalism for static solutions and derive the conserved charges. To the best of our knowledge, this formalism had not been previously derived for 5D black holes, including neutral ones. Finally, we show the emergence of Love symmetry in the near-zone regime, and derive the generators of the associated $sl(2,\mathbb{R})$ algebra. It is shown that all conditions for Love-number vanishing can be explained by this algebra in terms of the highest-weight property.

		\end{abstract}
		
		\pagebreak
		

		\section{Introduction}
		Love numbers are dimensionless parameters that quantify the linear response of an object perturbed by an external tidal potential. These quantities were first introduced in 1909 by  Augustus Edward Hough Love~\cite{Love}, who investigated  tidal deformations induced on Earth's surface in the framework of Newtonian gravity. This formalism has since been extended to the domain of general relativity~\cite{LNBH,LNNS, Flanagan, Hinderer, Poisson}, being used to investigate tidal deformations induced in objects such as neutron stars and Black holes. As well as their role in investigations concerning changes of shape and size of bodies affected by tidal perturbations, Love numbers are also relevant as means of probing the internal structure of massive compact objects and provide insight on the equation of state of such ultra-dense objects, which provide important information for astrophysical and cosmological models. Moreover, tidal deformations may vary between theories of gravity, so that Love numbers may be used as tools for testing General Relativity (GR) and constraining competing models. Given their nature as linear tidal responses, information about Love numbers can be extracted from gravitational wave detections~\cite{LIGOScientific:2016aoc, Maselli:2021men}, making this topic particularly relevant in the era of high-sensitivity gravitational wave observations. They are also important in point-particle effective field theory (EFT) descriptions of gravity, where these parameters are encoded in the EFT action as Wilson coefficients for quadratic operators in the curvature~\cite{EFT1, EFT2, EFT3, EFT4}. In this approach, Love number computation is reduced to a matching condition between the worldline EFT and full classical predictions.
		
		Love numbers can be obtained by means of a multipolar expansion of the perturbative wave equation, upon which the desired parameters emerge as a proportionality factor relating the induced mass multipole moment and the external tidal field~\cite{Poisson}. In practice, this procedure is greatly facilitated by the presence of certain symmetries, related to the existence of Killing or Killing-Yano tensors~\cite{Yano, FrolovI,FrolovII}, which allow for separability of the wave equation. We may thus use a mode decomposition to separate the wave equation into angular and radial parts. For asymptotically flat black hole spacetimes that are regular at the outer horizon, which we describe topologically as an $n$-sphere, the radial wave function behaves asymptotically as~\cite{PRD108}
		\begin{equation}
			R_{\ell}(r)\approx A_{\ell} r^{\ell} + B_{\ell}r^{1-\left(\ell+n\right)},
		\end{equation}
		in which case the response coefficients are identified with the ratio $B_{\ell}/A_{\ell}$. Love numbers correspond to the real part of these coefficients, which describes conservative tidal deformations, while the imaginary part is associated with dissipative effects. Here, ``infinity" is identified with the location of the tidal source, so that the growing mode proportional to $r^{\ell}$ is not considered unphysical.
		
		In four dimensional GR, static Love numbers can be shown to vanish~\cite{Poisson} on general grounds. Nevertheless, this need not be the case for Black holes in higher dimensions, which are in general nonzero ~\cite{PRD108, BHstereotyping ,Charalam_2023, Nair, Gray, LeTiec} even in the static limit. In this sense, higher-dimensional black holes provide valuable insight into the deeper aspects of Love number physics, as the nontrivial vanishing conditions allow for better understanding of the relationship between tidal responses and physical observables. In fact, gravity has been argued to be richer in dimensions higher than four~\cite{Emparan}, as uniqueness theorems for black holes become far less stringent in these settings, thus allowing for a richer landscape of black hole solutions. Reasons for this added richness include the higher number of independent angular momenta and the competition between gravitational and centrifugal effects, as the former are dimension dependent, while the latter are not~\cite{Emparan}. Moreover, the relevance of higher-dimensional black holes extends beyond gravitational physics, as the well-known AdS/CFT correspondence can be used to link the physics of these objects to that of strongly coupled quantum field theories in lower dimensions~\cite{Maldacena}. Another strong motivation for the study of higher-dimensional theories is derived from the observation that  gravitational interaction simplifies dramatically in the limit $D \to \infty$, and may allow for analytical treatment of otherwise complicated or intractable issues~\cite{Emparan2013, Emparan2020, Suzuki2015}. Thus, investigations of black holes in five dimensions provide an important initial step toward unified $D+1$ treatments of these solutions, wherein the $D \to \infty$ may be applied. On the experimental side, it should be mentioned that it has been recently argued that  the shadow diameter of supermassive black holes such as M87* can be used as a probe into higher-dimensional gravity, and as means to provide bounds into the size and physics of extra dimensions~\cite{Vagnozzi}.  
		
    	Outside of the static regime, the phyiscs of tidal deformations becomes more complex. Indeed, dynamical Love numbers, which correspond to modes with $\omega\neq 0$, can be nonzero even in the fourth-dimensional case~\cite{LeTiec, Perry_arx23}. Such effects can be physically explained by frame-dragging caused by the black hole's rotation. Thus, tidal coefficients also convey dynamical information about extended gravitational objects, and precise knowledge of their dependence on angular frequency is necessary for accurate interpretation of gravitational wave signals.

	The vanishing of Love numbers in static, four-dimensional case, as well as for dynamic perturbations in some special cases, can be understood in terms of a hidden ${SL}(2,\mathbb{R})$ - or ``Love" - Symmetry, which can be demonstrated to exist in the near-zone limit, valid for long-wavelength perturbations in the outer region of the black hole. It can be shown that the vanishing of static Love numbers in four-dimensional black holes follows from the highest-weight property applied to the irreducible ${SL}(2,\mathbb{R})$ representation implied by $SL(2,\mathbb{R})$~\cite{CharalamPRL2021} derived from the near-zone equation. This symmetry is also found in higher dimensions, despite the existence of nonzero static Love numbers in this scenario. This is explained by the fact that arbitrary solutions of the near-zone equation do not belong to the relevant highest-weight representation of this symmetry. On the other hand, those states that do belong to a highest-weight representation can still be shown to lack a conservative tidal response even in higher dimensions. Moreover, a deep connection between Love symmetry and holography has been suggested~\cite{CharalamPRL2021, Gray, LoveSymmetry, Guevara}. In fact, this structure can be linked to the  $\rm{SL}(2,\mathbb{R})\times\rm{U(1)}$ near-zone isometry which underpins the Kerr/CFT correspondence in the four-dimensional case~\cite{CharalamPRL2021,Guica}, and which has been subsequently extended to more general black holes in higher dimensions~\cite{Lu1, Lu2}. The holographic principle can be used as tool in the calculation of quasinormal modes, Love numbers, Greybody factors, etc.~\cite{Kehagias, Bonelli}, and provides a valuable tool for the probing of black hole structure at finer levels. In this framework, Love symmetry can be viewed as a statement about the response properties of the dual CFT.  Hidden symmetries may also explain why some results, such as the aforementioned vanishing of Love numbers, are robust against quantum corrections, potentially helping resolving naturalness problems~\cite{LoveSymmetry, CharalamPRL2021, Hui, Charalam2025}.

		Despite its remarkable predictive power, GR is not generally believed to be a final theory, which is why frameworks such as supergravity and string theory are often considered in frontier investigations. Black hole solutions are particularly important in the string-theoretical setting, where they can be described in terms of D-branes~\cite{Strominger, MaldacenaSusskind}. This work deals with five-dimensional solutions of the STU family of black holes~\cite{cy96b, Behrnd, STUunveiled} derived in the setting of minimal $\mathcal{N}=2$ supergravity coupled to three vector supermultiplets, through the solution generating technique detailed in Ref.~\cite{cy96a}. These are the generating solutions from which the most general charged rotating black holes of $\mathcal{N}=4$ superstring vacua conforming with no-hair theorem can be obtained through U-duality~\cite{UDuality}. Such black holes appear naturally in five-dimensional theories with supersymmetry, and their generality makes them particularly useful in that they contain many important solutions such as Kerr/Myers-Perry, as well as their charged variants, as limits achieved though the vanishing of the appropriate charge parameters. Another important motivation for the study of STU black holes comes from the known correspondence between these solutions and qubits~\cite{QubitI, QubitII, QubitIII}, which provides a powerful tool through which black hole theory can be used to boost discoveries in the field of quantum information, and vice-versa. Several other aspects of STU black hole theory have been investigated in the literature, and a few recent developments in the subject can be found in Refs.~\cite{Hristov, Mai, Separability, Chattopadhyay, Sekhmani, DiracSTU, Wu, Cvetic:2013roa, Cvetic:2014vsa, Cvetic:2016bxi, Cvetic:2017zde, Cvetic:2019ekr, CveticPRD106}.

		Love symmetry is not a general feature of string-theoretical black holes. Indeed, sufficient conditions for the emergence of Love symmetry in Type-II superstring theory black holes, have been investigated in~\cite{Charalam2024}, where the Callan-Myers-Perry and $\alpha'^3$-corrected  Schwarzschild-Tangherlini black holes are shown to be counterexamples. Thus, higher-order corrections from string theory may leave measurable signatures in tidal perturbations, which may one day be used to differentiate those solutions from each other and from GR black holes. On the other hand, there are some string-theoretical black holes, such as the $D=4$ STU solution~\cite{CveticPRD105}, where Love symmetry still emerges within the near-zone limit. In the aforementioned reference, it was shown that this symmetry implies the vanishing of static Love numbers, in a similar fashion to what was found for uncharged black holes.   As we shall demonstrate in this paper, Love symmetry is also present in the fifth-dimensional version of the theory, although, as is typically the case in higher dimensions, this symmetry structure does not generally preclude conservative tidal responses even in the static limit. Nevertheless, we demonstrate that the special cases in which Love number vanishing does occur can be explained in terms of highest-weight representations of the associated $sl(2,\mathbb{R})$ algebra, in a manner analogous to the four-dimensional case~\cite{CveticPRD105}. Understanding whether or not Love symmetry should be seen as a fundamental property of black hole solutions derived from string theory is important as means of understanding the role of this symmetry within a quantum description of gravity, and may potentially serve as a tool to be used in comparisons between competing models from string theory. 
		
		Another important tool in investigations concerning tidal deformations is the Ladder symmetry approach. It can be shown that the perturbative equations in curved black holes with certain symmetries lead to a Hamiltonian formulation which can be associated to raising and operators that connect the solutions in a manner analogous to the ladder structure encountered for the quantum harmonic oscillator. This approach can potentially simplify the tidal deformation analysis through the use of a conserved current that follows from Ladder Symmetry, and allows one to describe solutions in the $\ell$-th level of the Ladder in terms of the known properties of a seed solution, that is often a constant due to homogeneous symmetry. This approach has been successfully applied to the most important black holes in $4D$ general relativity, including Schwarschild, Reissner-Nordström, Kerr and other black holes with separable perturbation equations. See, for example, ~\cite{Hui,HuiII,Katagiri, Ghosh, Chia, LadderRN} for investigations of this kind. Despite the aforementioned success in the four-dimensional case, we do not know of any work that has extended these results to five-dimensional black holes. Due to the solution generation technique used in the derivation of the STU metric considered in the current paper, the Ladder structure we shall derive in this paper will also prove valid for special cases such as  Schwarzschild-Tangherlini, Myers-Perry, and their charged variants, as these operators can be written entirely in terms of the seed Myers-Perry parameters in the static case.

		To summarize, this work aims to calculate Love numbers for the three-charged rotating STU black hole in five-dimensional spacetime and investigate the emergence of Love symmetry in the appropriate limit. The paper is structured as follows: after this introduction, we start with a short review introducing the STU metric and the relevant thermodynamic quantities that will be used throughout the work. We then consider the massless Klein-Gordon equation and use separation of variables to write the angular and radial scalar wave equations. In the two following sections, we solve the radial equation in the static and near-zone regimes respectively, and thoroughly investigate several aspects of the black hole tidal response in these scenarios. The next section is devoted to investigation of the Love symmetry structure that emerges in the near-zone regime. We derive the associated algebra generators appropriate to the STU background and Casimir element, and show that the vanishing of Love numbers in special cases can be explained in terms of this symmetry. This discussion is followed by a concluding section, where our results are summarized and perspectives for future investigations are discussed.

		\cmmnt{With this in mind, we have previously studied the 
			electrodynamics
			\cite{Cvetic:2013roa}, the initial value problem \cite{Cvetic:2014vsa}, the
			photon sphere and sonic horizons  \cite{Cvetic:2016bxi}, the equatorial
			timelike geodesics \cite{Cvetic:2017zde}, and the
			stability of massless, minimally-coupled scalar fields \cite{Cvetic:2019ekr}, 
			in a remarkable
			six-parameter family of exact rotating black hole
			solutions  of ungauged supergravity theory known as \emph{STU
				black holes} \cite{cy96b}.  }

		\section{5D STU  Rotating Black Holes}
		In this section, we provide a brief review of the STU geometry we shall me concerned with, which corresponds to a family of black hole solutions parametrized by their mass, two independent angular momenta, and three $\rm{U(1)}$ charges. STU black holes are solutions of  $\mathcal{N}=2$ gauged supergravity coupled to three vector superfields, which is a consistent truncation of maximally supersymmetric ungauged supergravity theories.   \cmmnt{These are the generating solutions for the most general  black holes of  five-dimensional maximally supersymmetric ungauged supergravity theories, which can be obtained as the low energy effective
			action of the heterotic   and type II string theories, toroidally compactified to five dimensions, leading to $N=4$ and $N=8$ ungagued supergravity theories, respectively.}

		\label{sec:5DSTU}
		\subsection{The STU Metric}

		In this section, we present the metric and notation used throughout the paper. In five dimensions, the general no-hair black hole solution of $\mathcal{N}=4$ superstring vacua, which can have up to $27$ charges, is obtained by subjecting a generating solution to \(\bigl[SO(5)\times SO(21)\bigr] / \bigl[SO(4)\times SO(20)\bigr] \;\subset\; O(5,21)\) transformations~\cite{cy96a}. The thus found solutions are the most general black holes that can be obtained from the low energy effective action of toroidally compactified heterotic and type II string theories, which give rise respectively to $\mathcal{N}=4$ and $\mathcal{N}=8$ ungagued supergravity theories. In this work we shall deal with these generating solutions, which consist of a class of five-dimensional black holes parameterized by their mass $M$, two angular momenta $J_{R,L}$, and three  independent $U(1)$ charges $Q_i$. 
		
		The metric is obtained through the solution generation technique explained in~\cite{cy96a}, which consists in performing Kaluza-Klein reduction and then applying a subset of the symmetry transformations of the effective three-dimensional action on the neutral rotating black hole  (also known as the Myers-Perry solution), which serves as a seed. To obtain the generating solution of interest, three independent $SO(1,1)\subset O(8,24)$ boosts must be applied to the seed neutral black hole. Each of these boosts corresponds to a parameter $\delta_{i}$. As we said before, the full procedure is detailed in Ref.~\cite{cy96a}; here we shall be content with presenting the metric for the three-chaged solution, which can be cast in the following, compact form, first presented in~\cite{Chong:2006zx},
		\begin{equation}
			ds^2_5 = - \Delta^{-2/3}G ( dt+{\cal A})^2 + \Delta^{1/3} ds^2_4~,
		\end{equation}
		where
		\begin{equation}
			ds^2_4 = {dx^2\over 4X} + {dy^2\over 4Y} + 
			{U\over G} (d\chi - {Z\over U} d\sigma)^2 + {XY\over U} d\sigma^2~, \label{KKansatzmetric}
		\end{equation}
		with 
		\cmmnt{
			\begin{subequations}
				\bea
				X & = &(x+a^2) (x+b^2) - \mu x~, \\
				Y & = &- (a^2 - y) (b^2 - y)~,\\
				G & =& (x+y) (x+y - \mu )~, \\
				U & = &yX - xY~, \\
				Z & = &ab(X+Y)~,\eea
			\end{subequations}
			
			\bea
			\Delta & = &(x+y)^3 H_1 H_2 H_3~, \\
			H_i & =& 1 + {\mu \sinh^2\delta_i\over x+y}~,~(i=1,2,3)~,\\
			{\cal A} & =& \frac{\mu \Pi_c }{ x + y - \mu}[ (a^2 + b^2 - y)d\sigma - abd\chi]
			-{\mu  \Pi_s \over x+y} (abd\sigma - y d\chi)~,
			\eea
			
			\bea
			x&=&r^2~,\nonumber\\
			y&=&a^2 \cos^2\theta + b^2\sin^2\theta~,\nonumber\\
			\sigma & = &{1\over a^2- b^2} \left(a\phi
			-b\psi \right)~,\nonumber\\
			\chi & = &{1\over a^2- b^2} \left( b\phi
			-a\psi\right)~.
			\eea
		}
		\begin{subequations}
			\bea
			X & = &(x+l_1^2) (x+l_2^2) - \mu x~, \\
			Y & = &- (l_1^2 - y) (l_2^2 - y)~,\\
			G & =& (x+y) (x+y - \mu )~, \\
			U & = &yX - xY~, \\
			Z & = &l_1l_2(X+Y)~,\eea
		\end{subequations}
		where $l_1$,$l_2$ and $\mu\equiv 2m$ are parameters of the seed Perry-Myers black hole,  and remain unchanged under solution generating procedure. 
		
		On the other hand, the functions $\Delta (x,y)$ and ${\cal A}(x,y)$ which, in the Myers- Perry black hole are given respectively by $\Delta_{0}  = (x+y)^3$ and ${\cal A}_{0}  = \frac{\mu}{ x + y - \mu}[ (l_1^2 + l_2^2 - y)d\sigma - l_1l_2d\chi]$,
		are modified by the $SO(1,1)\subset O(8,24)$ boosts to
		\bea
		\Delta & = &(x+y)^3 H_1 H_2 H_3~, \\
		H_i & =& 1 + {\mu \sinh^2\delta_i\over x+y}~,~(i=1,2,3)~,\\
		{\cal A} & =& \frac{\mu \Pi_c }{ x + y - \mu}[ (l_1^2 + l_2^2 - y)d\sigma - l_1l_2d\chi]
		-{\mu  \Pi_s \over x+y} (l_1l_2d\sigma - y d\chi)~,
		\eea
		where $\Pi_c\equiv \Pi_{i=1}^3\cosh\delta_i$ and $\Pi_s\equiv \Pi_{i=1}^3\sinh\delta_i$.
		
		In the above, we have used the base space coordinates $(x,y,\sigma,\chi)$, which are 
		related to the usual system of coordinates by the transformations
		\begin{subequations}
			\begin{align}
				x&=r^2~,\\
				y&=l_1^2 \cos^2\theta + l_2^2\sin^2\theta~,\\
				\sigma &= {l_1\phi
					-l_2\psi \over l_1^2- l_2^2} ~,\\
				\text{and} \hspace{40pt} &	\nonumber\\
				\chi &= {l_2\phi
					-l_1\psi\over l_1^2- l_2^2} ~.
			\end{align}
		\end{subequations}

		%
		%
		In the above, we assume $l_2^2>l_1^2$ so that $y\in (l_1^2,l_2^2)$ as $\theta\in (0,\pi/2)$.



		
		\subsection{Conserved Charges and angular momenta}

		In this subsection, we introduce the ADM mass, three charges and two independent angular momenta. Together, these parameters completely specify any five-dimensional black hole compatible with the ``no-hair" theorem. These quantities can all be described by combinations involving the seed  parameters and $SO(1,1)$ boosts $\delta_i$, all of which reduce to the corresponding quantities of the neutral Myers-Perry solution when all three $\delta_i$ go to zero.
		
		The mass and charges of the generating solutions can be calculated by standard methods (see, for example,~\cite{cy96a,cl97a}) and are written in the
		form:\footnote{As remarked above, we choose to work with the parameter $\mu=2m$ where $m$ is the usual mass parameter used in black hole literature, and used in Ref.~\cite{cy96a} for the STU black hole. We use duality invariant units such that the five  
			dimensional gravitational constant is given by $G_5={\pi/ 4}$. 
		}: 
		\bea
		M &=& {1\over 2}\mu\sum_{i=1}^3 \cosh 2\delta_i~, \label{Mass} \\
		Q_i &=& {1\over 2}\mu \sinh 2\delta_i~~~;~~i=1,2,3~. \label{charges}
		\eea
		Saturation of the BPS (or Bogomol'nyi) bound corresponds to the simultaneous limits $\mu\rightarrow 0$ and 
		$\delta_i\rightarrow\infty$, with $Q_i$, $M$ and the angular momenta derived below kept fixed. Thus, $\mu$ can be seen as a measure of the deviation from BPS saturation. 
		
		The metric symmetries give rise to two conserved angular momenta, identified with rotations parameterized by the coordinates $\phi$ and $\psi$. From the point of view of string theory, it is more interesting to work with left and right moving contributions. Rotational symmetry allows for the definition of two independent angular momenta $J_{R}$ and $J_{L}$, which are given by
		\be
		J_{R,L}= {1\over 2}\mu (l_1\pm l_2)(\prod_i \cosh\delta_i \mp
		\prod_i \sinh\delta_i)~,
		\label{eq:l12def}
		\ee
		where $l_{1}$ and $l_2$ are angular momentum parameters of the seed Myers-Perry black hole used in the solution-generation procedure. The ``usual" angular momenta can be derived from these quantities through the combinations $J_{\phi}=J_R+J_L$, $J_{\psi}=J_R-J_L$, which are quantized as integers.
		
		\subsection{Thermodynamic Quantities}
		
		One of  the most important aspects in black hole theory is the description of their macroscopic thermodynamic properties, which is important not only for the deepening of our understanding of black holes, but also because of the expected connection between this area of investigation and the development of consistent quantum gravity models. A relevant feature of the STU model is the fact that black hole thermodynamics can be described in terms of left and right mode contributions, which suggests a microscopic interpretation of the entropy arising from the counting of left and right moving solitonic states in the underlying string model. We shall for these reasons point out some of the most important thermodynamic variables in this framework, and throughout the work it will be seen that many of the special features of the tidal perturbations we are concerned with have a simple explanation in terms of these quantities. The results of this section are those found in~\cite{cy96b} and~\cite{cy96a}. See also~\cite{cl97a, cl97b} for other relevant developments related to this subject.
		
		In agreement with the above remarks, the entropy of STU black holes at the outer (inner) horizons can be described as the sum (difference) of left and right moving contributions, namely,
		\be
		S_{\pm}=S_L\pm S_R\, ,
		\ee
		where
		
		\begin{equation}
			\label{eq:SL}
			\begin{split}
				S_L &= 2\pi \sqrt{{1\over 4}\mu^3  
					(\prod_i \cosh\delta_i+\prod_i \sinh\delta_i)^2-J^2_L} \\
				&= \pi\mu (\prod_i \cosh\delta_i+\prod_i \sinh\delta_i)
				\sqrt{\mu - (l_1-l_2)^2}~
			\end{split}
		\end{equation}
		\hspace{70pt} and
		\begin{equation}
			\label{eq:SR}
			\begin{split}
				S_R &= 2\pi  
				\sqrt{{1\over 4}\mu^3  
					(\prod_i \cosh\delta_i-\prod_i \sinh\delta_i)^2-J^2_R} \\
				&= \pi\mu (\prod_i \cosh\delta_i-\prod_i \sinh\delta_i)
				\sqrt{\mu - (l_1+l_2)^2}~.
			\end{split}
		\end{equation}
		\cmmnt{\bea
			S_L&=& 2\pi \sqrt{{1\over 4}\mu^3 
				(\prod_i \cosh\delta_i+\prod_i \sinh\delta_i)^2-J^2_L} \nonumber \\
			&=&\pi\mu (\prod_i \cosh\delta_i+\prod_i \sinh\delta_i)
			\sqrt{\mu - (l_1-l_2)^2}~\label{eq:SL} \\
			\text{and}  & \\
			S_R&=& 2\pi 
			\sqrt{{1\over 4}\mu^3 
				(\prod_i \cosh\delta_i-\prod_i \sinh\delta_i)^2-J^2_R} \nonumber \\
			&=&\pi\mu (\prod_i \cosh\delta_i-\prod_i \sinh\delta_i)
			\sqrt{\mu - (l_1+l_2)^2}~. \label{eq:SR}
			\eea}
		
		Similarly, the  inverse Hawking temperature of the Black hole at the inner and outer horizons can be written in the form:
		\be\label{inversetemperatures}
		\beta_{\pm}=\frac{\beta_L\pm \beta_R}{2}, 
		\ee
		where
		\begin{equation}
			\beta_L=
			{\pi\mu^2 (\prod_i \cosh^2\delta_i-\prod_i \sinh^2\delta_i)
				\over
				\sqrt{{1\over 4}\mu^3 (\prod_i \cosh\delta_i+\prod_i \sinh\delta_i)^2-J^2_L}}
			={2\pi\mu (\prod_i \cosh\delta_i-\prod_i \sinh\delta_i)\over
				\sqrt{\mu - (l_1-l_2)^2}}~,
			\label{eq:betal}
		\end{equation} 
		and
		\begin{equation}
			\beta_R =
			{\pi\mu^2 (\prod_i \cosh^2\delta_i-\prod_i \sinh^2\delta_i)
				\over\sqrt{{1\over 4}\mu^3 
					(\prod_i \cosh\delta_i-\prod_i \sinh\delta_i)^2-J^2_R}}
			={2\pi\mu (\prod_i \cosh\delta_i+\prod_i \sinh\delta_i)\over
				\sqrt{\mu - (l_1+l_2)^2}}~.
			\label{eq:betar}
		\end{equation}
		\cmmnt{
			\bea
			\beta_L&=& 
			{\pi\mu^2 (\prod_i \cosh^2\delta_i-\prod_i \sinh^2\delta_i)
				\over
				\sqrt{{1\over 4}\mu^3 (\prod_i \cosh\delta_i+\prod_i \sinh\delta_i)^2-J^2_L}}
			={2\pi\mu (\prod_i \cosh\delta_i-\prod_i \sinh\delta_i)\over
				\sqrt{\mu - (l_1-l_2)^2}}~,
			\label{eq:betal}
			\\ \text{and}  & \\
			\beta_R &=& 
			{\pi\mu^2 (\prod_i \cosh^2\delta_i-\prod_i \sinh^2\delta_i)
				\over\sqrt{{1\over 4}\mu^3 
					(\prod_i \cosh\delta_i-\prod_i \sinh\delta_i)^2-J^2_R}}
			={2\pi\mu (\prod_i \cosh\delta_i+\prod_i \sinh\delta_i)\over
				\sqrt{\mu - (l_1+l_2)^2}}~.
			\label{eq:betar}
			\eea
		}
		We may use these formulae to define the surface acceleration at the outer and inner horizons, which will prove specially relevant in the following calculations. They are given by
			\begin{equation}\label{surfaceaccelerations}
			\kappa_{\pm}=\frac{4\pi}{\beta_{R}\pm \beta_L},
		\end{equation} 
		being thus proportional to the horizon temperatures. The angular velocities as
		defined at the outer  and inner horizon are
		\bea
		\Omega^R_{\pm} &=& {1\over 2}\left({d(\phi+\psi)\over dt}\right)_{ \pm} 
		\label{eq:rotvelR}~,\\
		\Omega^L_{\pm} &=& {1\over 2}\left({d(\phi-\psi)\over dt}\right)_{\pm} 
		\label{eq:rotvelL}~.
		\eea
		The rotational velocities and accelerations at the inner and outer horizons are not all independent thermodynamic quantities, being instead related according to
		${1\over\kappa_{-}}\Omega^R_{-}={1\over\kappa_{+}}\Omega^R_+$ 
		and $-{1\over\kappa_{-}}\Omega^L_{-}={1\over\kappa_{+}}\Omega^L_+$.
		The following important relations emerge:
		\bea
		\beta_+\Omega^L _+&=& {2\pi J_L\over \sqrt{{1\over 4}
				\mu^3 (\prod_i \cosh\delta_i+\prod_i \sinh\delta_i)^2-J^2_L}}
		={2\pi (l_1-l_2)\over\sqrt{\mu-(l_1-l_2)^2}}~,
		\label{eq:omegal}
		\\
		\beta_+\Omega^R_+ &=& {2\pi J_R\over \sqrt{{1\over 4}
				\mu^3 (\prod_i \cosh\delta_i-\prod_i \sinh\delta_i)^2-J^2_R}}
		={2\pi (l_1+l_2)\over\sqrt{\mu-(l_1+l_2)^2}}~,
		\label{eq:omegar}
		\eea
		which show that, although both the horizon temperatures and angular velocities are explicitly dependent on the boosts $\delta_i$, the above products can be written solely in terms of the seed Myers-Perry parameters $\mu, l_1$ and $l_2$. This means that these combinations are the same for all members of the STU family generated by a given seed solution. 
		\cmmnt{
			The $U(1)$ potentials for general rotating black holes are:
			\bea
			\beta_+\Phi^j_{L_+} &=& {\pi\mu
				(\tanh\delta_j 
				\prod_i\cosh\delta_i -
				\coth\delta_j \prod_i\sinh\delta_i)\over
				\sqrt{\mu - (l_1-l_2)^2}}~,
			\label{eq:U1L}\\
			\beta_+\Phi^j_{R_+} &=& {\pi\mu
				(\tanh\delta_j \prod_i\cosh\delta_i+\coth\delta_j \prod_i\sinh\delta_i)\over
				\sqrt{\mu - (l_1+l_2)^2}}~.
			\label{eq:U1R}
			\eea
		}

		\section{Minimally Coupled Massless Scalar Wave Equation}
		\label{sec:waveeq}
		Massless, minimally coupled scalar fields must satisfy the Klein-Gordon equation:
		\be
		{1\over\sqrt{-g}}\partial_\mu (\sqrt{-g}g^{\mu\nu}\partial_\nu \Phi) =0~.
		\ee
		Due to symmetries of the STU metric, the above equation is separable in this background. As shown in Ref.~\cite{cl97a}, solutions of the wave equation can thus be decomposed in modes of the form\footnote{Here, $r$ 
			is a five-dimensional
			analogue of the usual Boyer-Lindquist radial coordinate. It reduces to the 
			Schwarzschild-Tangherlini coordinate when charges and angular momenta vanish.}
		\be
		\Phi(t,r,\theta,\phi,\psi)\equiv\Phi_0(r)~\chi(\theta)~
		e^{-i\omega t+im_\phi\phi+im_\psi\psi}
		=\Phi_0(r)~\chi(\theta)~e^{-i\omega t+im_R(\phi+\psi)+im_L(\phi-\psi)}~.
		\ee
		The angular variables $\phi$ and $\psi$ have period $2\pi$;
		so $m_{\phi,\psi}=m_R\pm m_L$ are integer valued. In general, the frequency $\omega$ is complex-valued. However, mode stability for the five-dimensional STU metric has been demonstrated in~\cite{CveticPRD106}, which implies that modes with $\rm{Im}(\omega)>0$ are ruled out. Frequencies belonging to the lower half of the complex plane are still allowed, but these represent perturbations that vanish exponentially fast. While important in some applications, these perturbations are not particularly relevant to our current purposes, as the detection of tidal perturbations will be dominated by the oscillatory terms, which represent a persistent physical field, rather than transient effects. Thus, we may assume $\omega\in\mathbb{R}$ when interpreting our results. That said, we will see that the near-zone and static solutions are given in terms of hypergeometric functions, which are well-defined in most of the complex plane, so that the distinction between real and complex $\omega$ is only important in the process of extracting the real and imaginary parts of the response coefficients. 
		
		In our calculations, it will prove useful to introduce a dimensionless radial coordinate $\bx$ related 
		to $r$ by the transformation:
		\be
		\bx \equiv {r^2 - {1\over 2}(r^2_{+}+r^2_{-})\over
			(r^2_{+}-r^2_{-})}~,
		\label{eq:xdef}
		\ee
		where $r_{\pm}^2$ are the solutions of the horizon equation:
		\be
		(r^2+l_1^2)\, (r^2+l_2^2) -\mu r^2=0\, ,
		\ee
		and thus:
		\be\label{Defrprm}
		r_{\pm}^2 = \frac{1}{2}(\mu -l_1^2-l_2^2 \pm\Delta)\, ,
		\ee
		where
		\be\label{Delta}
		\Delta=\sqrt{[\mu-(l_1-l_2)^2][\mu-(l_1+l_2)^2]}~, 
		\ee
		which is a real quantity as long as the constraint $\mu\geq (|l_1|+|l_2|)^2$ is imposed. Note that the inner and outer horizons are written exclusively in terms of the parameters $\mu$, $l_1$ and $l_2$ of the seed Myers-Perry black hole, with no dependence on the boosts. Thus,  horizon positions are the same for all members of the family generated by a given seed, regardless of charges. 
		
		In terms of the coordinate $\bx$, the outer and inner horizons $r_{+}$ and $r_{-}$ 
		are located at $\bx={1\over 2}$ and $\bx=-{1\over 2}$, respectively, while the 
		asymptotically flat 
		region is identified with the limit $\bx\to\infty$. With this notation the wave equation can 
		be written as:
		\begin{align}
			{\partial\over\partial \bx}\left(\bx^2-{1\over 4}\right){\partial\over\partial \bx}\Phi_0(\bx)
			&+{1\over 4}\left[\bx\Delta\omega^2-\Lambda+M\omega^2-{1\over \bx+{1\over 2}}\left({\omega\over\kappa_{-}}+2m_R {\Omega^R_+\over\kappa_{+}}
			-2m_L {\Omega^L_+ \over\kappa_{+}}\right)^2\right. 
			\label{eq:geneq}
			\\
			&\left.{1\over \bx-{1\over 2}}
			\left({\omega\over\kappa_{+}}-2m_R {\Omega^R_+\over\kappa_{+}}
			-2m_L {\Omega^L_+\over\kappa_{+}}\right)^2
			\right]\Phi_0(\bx) = 0~. \nonumber
		\end{align}

		Here, $\kappa_{\pm}$ and $\Omega^{R,L}_{\pm}$ are as defined in the previous section, $M$ is the mass, and $\Lambda$ is the 
		eigenvalue of the angular Laplacian:
		\begin{equation}
			\hat{\Lambda}=
			\vec{K}^2+(l^2_1+l_2^2)\omega^2+(l^2_2-l_1^2)\omega^2\cos 2\theta~,
			\label{eq:lambdadef}
		\end{equation}
		
		where
		\be
		\vec{K}^2 =
		- {1\over \sin 2\theta}{\partial\over\partial\theta}
		\sin 2\theta{\partial\over\partial\theta}
		-{1\over \sin^2 \theta}{\partial^2\over\partial\phi^2}
		-{1\over \cos^2 \theta}{\partial^2\over\partial\psi^2}~,
		\label{eq:flatlap}
		\ee
		is the usual $3$-sphere Laplacian, to which $\hat\Lambda$ is reduced in the static case. Note that the only remnants of the solution generation technique are the boosts $\delta_i$ which only appear in the above scalar equation through the thermodynamic quantities $\kappa_{\pm}$ and $\Omega^{R,L}_{\pm}$. Particular cases such as neutral black holes or solutions with fewer charges can be described by~\eqref{eq:geneq} as long as one substitutes the corresponding angular accelerations and velocities.

		

		\section{Static solutions and tidal responses}
		\label{sec:solutions}
		
		The scalar equation governing time-independent modes is obtained by setting $\omega=0$ in the radial equation~\eqref{eq:geneq} and noting that $\hat{\Lambda}$ reduces to $\vec{K}^2$ in this case. The angular equation can thus be solved in terms of fifth-dimensional spherical harmonics (see, for example, Ref.~\cite{SphericalHarmonics} for a review on the subject), whereas the radial equation simplifies to
		\be
		{\partial\over\partial \bx}\left(\bd(\bx){\partial\over\partial \bx}\Phi_{\ell }(\bx)\right)
		\, +\left[- \frac{\Lambda_{\ell}}{4}+\frac{\left(m_R {\Omega^R_+\over\kappa_{+}}+m_L {\Omega^L_+\over\kappa_{+}}\right)^2}{ \bx-{1\over 2}}
		-\frac{\left(m_R {\Omega^R_+\over\kappa_{+}}
			-m_L {\Omega^L_+\over\kappa_{+}}\right)^2}{ \bx+{1\over 2}}\right] \Phi_{\ell }(\bx) = 0, \label{eq.str}\ee
		where 
		\begin{equation}\label{Def:BarDelta}
			\bd(\bx)\equiv\bx^2-\frac{1}{4}
		\end{equation} and $\Lambda_\ell\equiv{\ell }({\ell }+2)$ is the 
		eigenvalue of the $3$-sphere Laplacian ${\hat \Lambda} =\vec{K}^2 $ defined in Eq. \eqref{eq:flatlap}. We see that this equation has three singularities located at $\bx=\pm 1/2$ and infinity. All of these singular points can be verified to be regular, so that the above equation must be reducible to the hypergeometric form through an appropriate transformation, which we perform explicitly in the next subsection.  
		
		The massless time-independent Klein-Gordon equation in the Myers-Perry background is of course a limiting case of~\eqref{eq.str}, achieved when the three boost parameters $\delta_i$ ($i=1,2,3$) are set to zero. Note, however, that the complexity of the static equation remains the same as it was for neutral black holes. In fact this equation, and thus the static solutions, can be written entirely in terms of the seed Myers-Perry parameters $\mu$, $l_1$ and $l_2$. This is possible because, due to Eqs.\eqref{eq:omegal}, \eqref{eq:omegar} and the definitions of $\k_{\pm}$, we can derive the relations

		\be
		\kappa^{-1}_{+}\Omega^L_+={ (l_1-l_2)\over\sqrt{\mu-(l_1-l_2)^2}}~,\ \ \ 
		\kappa^{-1}_+\Omega^R_+ 
		={(l_1+l_2)\over\sqrt{\mu-(l_1+l_2)^2}}~. \label{eq.ok}
		\ee
		Thus the solution depends only on the {\it  seed parameters of the five-dimensional neutral rotating black hole}! 
		This is an intriguing observation that can be traced to the solution generation technique used to obtain the STU black hole metric.
		Namely, one can show explicitly that  full  scalar Laplacian  {for static scalar field does not depend on  $\Delta(x,y) $ and 
			${\cal A}(x,y)$}  in the metric Ansatz~\eqref{KKansatzmetric}. We should point out that, since this structure is a consequence of the solution generating technique, it is thus applicable to other  black holes solutions of the Lagrangians that have a five-dimensional Myers-Perry solution as a special case.

		\subsection{Raising and Lowering Ladder Operators}
		
		One  important feature of Eq.~\eqref{eq.str} is that it is solved by hypergeometric functions, as we shall now demonstrate. To this end, we introduce the variable 
		\begin{equation}
			z= \frac{\bx-\frac{1}{2}}{\bx+\frac{1}{2}},
		\end{equation}
		with use of which one can write the time-independent radial Klein-Gordon equation in the form
		\begin{equation}
			\left(z(1-z)\frac{d^2}{dz^2} + (1-z)\frac{d}{dz} + \frac{(a_R + a_L)^2}{z}-\frac{l(l+2)}{4(1-z)}-(a_R - a_L)^2\right)\Phi_{l}(z)=0,
		\end{equation}
		where we have defined
		\be
		a_R\equiv m_R\frac{\Omega^R_+}{\kappa_+}=m_R{(l_1+l_2)\over\sqrt{\mu-(l_1+l_2)^2}} \ \ \ \text{and} \ \ \  \   a_L\equiv m_L\frac{\Omega^L_+}{\kappa_+}=m_L{ (l_1-l_2)\over\sqrt{\mu-(l_1-l_2)^2}}.
		\label{eq.aRaL}\ee
		
		The above equation is easily solved with use of an ansatz $z^A(1-z)^Bf(z)$ which leads to a standard Hypergeometric equation~\cite{NISTI}. After returning to the original variable $\bx$, the static scalar solution is found to be
		\bea
		\Phi_{\ell } (\bx) &=&{\cal A}_\ell \, 
		\left({\bx-{1\over 2}\over \bx+{1\over 2}}\right)^{{ i}(a_L+a_R)}\psi_{\ell}(\bx)\, , \ \ \ \ \text{where} \ \nonumber\\
		\psi_{\ell}(\bx)&=&\left(\bx+{1\over 2}\right)^{-\xi_{\ell}} \, 
		{_{2}F_{1}}\left(\xi_{\ell}+{2 {i}} a_{R},
		\xi_{\ell}+2{ i} a_L;
		1+2{ i}\left( a_L +a_R\right)  ;{\bx-{1\over 2}\over \bx+{1\over 2}}\right)~,
		\label{eq:phiin0}
		\eea
		where we have defined the parameters
		\begin{equation}
			\xi_{\ell}\equiv \frac{1}{2}(1+\sqrt{1+\Lambda_\ell})=1+\frac{\ell}{2} .
		\end{equation}
		and ${\cal A}_\ell$ is a normalization factor.
		
		We shall now develop a ladder formalism by deriving the appropriate raising and lowering operators, as well as the associated Hamiltonian. To this end, we first note the contiguous relation
		\begin{equation}\label{contiguous}
			(1-z)\,F(a+1,b+1;c;z)=F(a,b;c;z) +\frac{z}{c}\left(a+b-c+1\right)F(a+1,b+1;c+1;z)
		\end{equation}
		which one can easily verify by subjecting Eq. $(2.24)$ of Ref.~\cite{R} to the transformations $a\to a+1$, $b\to b+1$. We can combine~\eqref{contiguous} with the well-known differentiation formula~\cite{NISTI}
		\begin{equation}
			\frac{dF(a,b;c;z)}{dz}=\frac{ab}{c}F(a+1,b+1;c+1;z)
		\end{equation}
		to derive the relation
		\be
		(1-z)\,F(a+1,b+1;c;z)=F(a,b;c;z) +\alpha z\frac{dF(a,b;c;z) }{dz}\, , \label{eq:gcr}
		\ee
		where 
		\be
		\alpha =\frac{(a+b+1-c)}{ab}\, .
		\ee
		For the hypergeometric function in eq.~\eqref{eq:phiin0} we have, for each fixed choice of $\ell$,
		\begin{align}
			a &=\xi_{\ell}+2{ i} a_R, \\
			b &=\xi_{\ell}+2{ i} a_L \ \ \ \ \text{and}\\
			c &=1+2{ i}\left( a_L + a_R\right),
		\end{align}
		hence
		\be
		\alpha_{\ell}=\frac{2\xi_{\ell}}{ab}=\frac{2+\ell}{\left(1+\frac{\ell}{2}+2{ i} a_R\right)\left(1+\frac{\ell}{2}+2{ i} a_L\right)}.
		\ee
		
		We can now employ~\eqref{eq:gcr} to obtain the following relations between the $\psi_{\ell}(\bx)$:
		\be 
		\psi_{\ell +2}(\bx) = -\alpha_{{\ell}} \left[D^+_\ell\psi_\ell(\bx)\right]\, ,
		\ee
		with a raising operator 
		\bea
		D^ +_\ell &=& -\left(\bx^2-\frac{1}{4}\right)\partial_{\bx} -\xi_{{\ell}}\left(\bx-\frac{1}{2}\right) -\frac{1}{\alpha_{{\ell}}}\nonumber\\
		&=& -\bd\partial_{\bx} -\xi_{\ell}\, \bx -{i} (a_L+a_R)+\frac{2a_La_R}{\xi_{\ell}}\, , \label{eq:raising}
		\eea
		\cmmnt{where, aiming for notational simplicity, we  omit here and henceforth the $l$ index from $\xi_{l}$, $\alpha_l$ etc., treating $l$ as fixed for the moment. We shall maintain this convention in the following derivations unless ambiguity arises.} Note that the prefactor  $-\alpha_{\ell}$ can be absorbed into the normalization factor  ${\cal A}_\ell$. Namely, one can choose ${{\cal A}_{\ell+2}}/{{\cal A}_{\ell}}=-\alpha_{\ell}$. 
		
		Under some particular conditions on the parameters and quantum numbers of the theory, there emerges a special ladder structure, resembling that encountered in standard $4D$ black holes. The solutions that fall in this category are of particular importance for, as shall be shown shortly, these are precisely the requirements that must be satisfied by black holes with vanishing static Love numbers. There are two such sets of conditions, namely,
		\begin{subequations}\label{StaticLNvanishingcond}
			\begin{align}
				&\ell= 2n ; \ n\in\mathbb{N} \hspace{30pt} \text{and} \hspace{30pt} a_R=0 \\
				\text{or} \hspace{30pt } &\nonumber \\
				&\ell= 2n ; \ n\in\mathbb{N} \hspace{30pt} \text{and} \hspace{30pt} a_L=0.
			\end{align}
		\end{subequations}
		
			Note that these conditions can only be realized for \emph{even} choices of $\ell$. Although the operators obtained from substitution of $a_{L/R}=0$ are of course also valid for odd $\ell$, the ensuing solutions will not be analogous to those of the Kerr black hole, nor will static Love numbers vanish if $\ell=2n+1$. Conditions~\eqref{StaticLNvanishingcond} impose restrictions on the particular quantum numbers of the wavefunction and/or the black hole parameters. As seen from the definitions~\eqref{eq.aRaL}, the conditions $a_{L/R}=0$are satisfied if either $m_{L/R}=0$ or $|l_1|=|l_2|$. Thus, these constraints represent limiting cases in which $(i)$ the ``azimuthal" quantum numbers are such that $m_{\phi}\pm m_{\psi}$ or $(ii)$ the independent angular momenta $J_{\phi}$ and $J_{\psi}$ have equal magnitudes.  In these cases, the raising operators simplify to
		\be\label{raising Kerr}
		D^+_\ell= -\bd\partial_{\bx} -\xi_{\ell} \bx -{{i}} {a_{R/L}}\,
		\ee
		which, as remarked, is of the same form as the analogue raising operator Kerr black hole theory, with the important difference that it now raises the solutions by \emph{two} steps, so that this analogue Kerr structure is generated among solutions with even $\ell$. In fact, once one makes the identification $\frac{\ell}{2}\leftrightarrow \tilde\ell$, with $\tilde\ell$ the parameters from the Kerr solution, it is seen that~\eqref{eq.str} itself shares this analogy when~\eqref{StaticLNvanishingcond} are satisfied.
		
		To derive the relevant lowering operators, we use the relation~\cite{R} 
		\begin{equation}\label{Rel130}
			\left(a-1\right)F(a,b;c;z)=\left(a+b-c-1\right)F(a-1,b;c;z)+\left(c-b\right)F(a-1,b-1;c;z),
		\end{equation}
		combined with the identity~\cite{NISTI}
		\begin{equation}
			\left(c-a\right)F(a-1,b;c;z)=\left(1-z\right)z\frac{dF(a,b;c;z)}{dz}-\left(a-c+bz\right)F(a,b;c;z)
		\end{equation}
		which follows from one of Gauss's contiguous relations. Substitution of the above equation into~\eqref{Rel130} leads to
		\begin{equation}\label{Rel_lowering}
			F(a-1,b-1;c;z)=-\beta\left\{ z(1-z)\frac{dF(a,b;c;z) }{dz}-\large[\frac{1}{\beta}(1-z)+z(a+b-c)]F(a,b;c;z)\right\}
		\end{equation}
		\cmmnt{
			\bea
			&&F(a-1,b-1,c,z)=\\
			&& -\beta\large\{ z(1-z)\frac{dF(a,b;c;z) }{dz}-\large[\frac{1}{\beta}(1-z)+z(a+b-c)]F(a,b;c;z) \large\}\, , \nonumber \label{eq:gcl}
			\eea}
		where
		\be\ \beta=\frac{(a+b-c-1)}{(c-a)(c-b)} \,.
		\ee
		For the parameters corresponding to the solution~\eqref{eq:phiin0} at level $\ell$, we have
		
		\be
		\beta_{\ell} =\frac{2(\xi_{\ell}-1)}{[(\xi_{\ell}-1)-2{ {i}}a_L][(\xi_{\ell}-1)-2{ {i}}a_R]}=\frac{\ell}{\left(\frac{\ell}{2}-2{ {i}}a_L\right)\left(\frac{\ell}{2}-2{ {i}}a_R\right)}
		\ee
		We can now employ~\eqref{eq:gcr} 
		to obtain the following relations among  $\psi_{\ell-2}(x)$ and $\psi_\ell(x)$:
		\be 
		\psi_{\ell -2}(\bx) = -\beta_{\ell} \left[D^-_\ell\psi_\ell(\bx)\right]\, ,
		\ee
		where the lowering operator is
		\bea
		D^-_\ell &=& (\bx^2-\frac{1}{4})\partial_{\bx} -(a+b-c-\xi_{\ell}) (\bx-\frac{1}{2})-\frac{1}{\beta_{\ell}} \nonumber\\
		&=& \bd\partial_{\bx} -(\xi_{\ell}-1)\bx
		+{i}(a_L+a_R)+\frac{2a_L a_R}{(\xi_{\ell}-1)}\, , \label{eq:lowering}
		\eea
		and we note the symmetric structure of  $D^{\pm}_\ell$. Similarly to what we did for the raising operator, we now turn our attention to the special cases $a_{L/R}=0$, where the lowering operator is reduced to:
		\be
		D^-_\ell= \bd\partial_{\bx} -(\xi_{\ell}-1){\bx} +{ i} a_{R/L}.
		\ee
		
		The ladder operators we just derived can be used to construct a Hamiltonian operator for the static Klein-Gordon equation. To do this, we write $\Phi_{\ell}(\bx)=g(\bx)\psi_{\ell}(\bx)$, where the prefactor $g(\bx)$, which can be directlty read off from Eq.~\eqref{eq:phiin0} satisfies
		\begin{equation}
			g_{\ell}(\bx)=A_{\ell}\left(\frac{\bx-1/2}{\bx+1/2}\right)^{{i}\left(a_L+a_R\right)}\implies \partial_{\bx}g_{\ell}(\bx)={i}\frac{\left(a_L+a_R\right)}{\bx^2-\frac{1}{4}}g_{\ell}(\bx).
		\end{equation}
		Using the above, we can write
		\begin{equation}
			\partial_{\bx}\left(\bd\partial_{\bx}\Phi_{\ell}(\bx)\right)=g_{\ell}(\bx)\left\{\partial_{\bx}\left(\bd\partial_{\bx}\psi_{\ell}(\bx)\right) +2{i}\left(a_L+a_R\right)\pb{\bx}\psi_{\ell}(\bx)-\frac{\left(a_L+a_R\right)^2}{\bd}\psi_{\ell}(\bx)\right\},
		\end{equation}
		Working with $\bx$ in the interval $\left(\frac{1}{2},\infty\right)$, we can hence write the radial scalar equation in the form 
		
		\begin{equation}\label{scalareq_psi}
			\left[\partial_{\bx}\bd\partial_{\bx} +2{i}\left(a_L+a_R\right)\pb{\bx}-\frac{1}{4}\ell(\ell+2) +\frac{4a_La_R}{\bx+\frac{1}{2}}\right]\psi_{\ell}(\bx)=0.
		\end{equation}
		
		Now we may use the fact that $\xi_{\ell + 2}-1=\xi_{\ell}$ to write
		\begin{equation}
			\begin{split}
				D_{\ell+2}^{-}D_{\ell}^{+}=&-\left(\bd\pb{\bx}-\xi_{\ell}\bx+\mathcal{Z}^*\right)\left(\bd\pb{\bx}+\xi_{\ell}\bx-\mathcal{Z}\right)=\\
				&-\bd\left[\pb{\bx}\bd\pb{\bx}+\xi_{\ell}+\left(\mathcal{Z}^{*}-\mathcal{Z}\right)\pb{\bx} + \left(\frac{\xi_{\ell}}{2}\right)^2-\xi_{{\ell}}\left(\mathcal{Z}+\mathcal{Z}^* \right)\bx+ |\mathcal{Z}|^2\right],
			\end{split}
		\end{equation}
		where $\mathcal{Z}\equiv \frac{2a_La_R}{\xi_{\ell}}-{i}\left(a_L+a_R\right)$. Since $\mathcal{Z}+\mathcal{Z}^*=\frac{4a_La_R}{\xi_{\ell}}$ and $\mathcal{Z}-\mathcal{Z}^*=-2{i}\left(a_L+a_R\right)$, we can write Eq.~\eqref{scalareq_psi} in the simple form
		\begin{equation}
			D_{\ell+2}^{-}D_{\ell}^{+}\psi_{\ell}=\frac{\psi_{\ell}}{\alpha_{{\ell}}\beta_{\ell+2}},
		\end{equation}
		which is equivalent to the Hamiltonian equation	
		\be\label{HamiltonianEq}
		{\cal H}_\ell \psi_\ell(\bx)=0\, ,
		\ee
		\cmmnt{where 
			{
				\be {\cal H}_\ell=
				-(\bx^2-\frac{1}{4})\left[ \partial_{\bx}\left(\bx^2-\frac{1}{4}\right) \partial_{\bx} +{ i}(a_L+a_R)\partial_{\bx}-\frac{1}{4}\ell(\ell+2) +\frac{a_La_R}{\bx+\frac{1}{2}}\right]\, .
				\ee
				Furthermore:}}
		where
		\begin{equation}\label{Hamiltonian}
			\begin{split}
				{\cal H}_\ell &=-\bd\left\{\partial_{\bx}\bd\partial_{\bx} +2{i}\left(a_L+a_R\right)\pb{\bx}-\frac{1}{4}\ell(\ell+2) +\frac{4a_La_R}{\bx+\frac{1}{2}}\right\}\\
				&= D^-_{\ell+2}D^+_{\ell}-\left\{\left(\frac{\xi_{\ell}}{2}\right)^2+a_L^2+a_R^2+\left(\frac{2a_La_R}{\xi_{{\ell}}}\right)^2\right\}.
			\end{split}
		\end{equation}

		We can use the Ladder symmetry we have derived to investigate the asymptotic behavior of solutions and the conditions that lead to vanishing Love numbers. We can investigate the tidal response by examining the Hamiltonian equation~\eqref{HamiltonianEq} for large $\bx$. In that regime, the last term in~\eqref{Hamiltonian} is of higher-order, and we must solve the asymptotic equation
		\begin{equation}\label{asymptEq}
			\bar{x}^2 \partial_{\bx}^2\psi_\ell(\bar{x}) + 2[\bx+{i}(a_L+a_R)]\partial_{\bx}\psi_\ell(\bar{x}) - \frac{\ell(\ell+2)}{4}\psi_\ell(\bar{x}) = 0,
		\end{equation}
		Note that, because the term proportional to $a_La_R$ becomes negligible at sufficiently large distances, Eq.~\eqref{asymptEq} has the same form (up to a coefficient change, of course) whether or not $a_L$ and $a_R$ are both nonzero.  The asymptotic equation implies the indicial polynomial $s^2 + s - \frac{\ell(\ell+2)}{4} = 0$, whose roots are
		\begin{align}
			s_1 = \frac{\ell}{2} \quad \text{and} \quad s_2 = -\frac{\ell+2}{2}.
		\end{align}

		These two coefficients give rise to two linearly independent solutions of~\eqref{asymptEq} of which the general solution is a linear combination, written in the form
		\begin{equation}\label{asymptsol}
			\psi_\ell(\bar{x}) \sim \  a_{\ell} \bar{x}^{\ell/2} \left[1+\frac{{i}(a_L+a_R)}{\bar{x}} \right] +  \frac{b_{\ell}}{\bar{x}^{(\ell+2)/2}}\left[1+ \frac{{i}(a_L+a_R)}{\bar{x}} \right], 
		\end{equation}
		which is precisely the $A_lr^{\ell} + B_{\ell}r^{-\ell-2} $ asymptotic behavior expected. We thus find that $b_{\ell}\neq 0$ is a necessary condition for a nonvanishing conservative response. In order to obtain $a_{\ell}$ and  $b_{\ell}$, and thus derive these responses explicitly, one needs to perform a careful analysis of the asymptotic properties of the solutions. This will be done in the next subsection. For our current purpose of investigating the relationship between Love number vanishing and the Ladder structure we have derived, it shall be sufficient to investigate the qualitative properties of these solutions, which can be easily derived from the Hamiltonian equation and Ladder operators.

		We can now use the Ladder structure to examine the conditions that lead to vanishing Love numbers. We do this in a manner analogous to that previously used in four-dimensional black holes. We first note that, for $\ell=0$ and either $a_R$ or $a_L$ equal to zero, the above Hamiltonian equation has a solution of the form
		\begin{equation}\label{homogeneous}
			\psi_{0}=a_0,
		\end{equation}
		which satisfies the required boundary conditions at infinity and is regular at the horizon, being thus an acceptable physical solution of the problem. Note that~\eqref{homogeneous} is simply a homogeneous (i.e., independent of $r$) solution. Moreover, the other linearly-independent solution, namely the one with asymptotic behavior $b_0/\bx$, can be shown to be irregular at the horizon as in the Kerr case (see, for example, ~\cite{Hui}). The action of the raising operator~\eqref{raising Kerr} on a solution with this asymptotic behavior gives
		\begin{equation}
			\psi_{2}\sim -\alpha_{0}a_{0}\left(\bx -{i}a_{L}\right).
		\end{equation}
		From the form of~\eqref{raising Kerr}, it is easily seem that the next application of this operator results in a second-degree polynomial, and repetition of this process leads to a tower of solutions $\psi_{2n}\propto (D_{\ell}^{+})^{n}\psi_{0}$, with $n\in\mathbb{N}$, which behave asymptotically as $n$-th degree polynomials. Going back from $\bx$ to the variable $r$, we thus find that, when $a_{R/L}=0$, the leading-order behavior of the radial solutions at even levels is of the typical quasi-polynomial behavior that is known to signal Love number vanishing in black hole theory~\cite{LoveSymmetry}. The fact that $D_{\ell}^{+}$ always raises $\ell$ by two clearly implies that solutions with odd $\ell$ cannot belong to the tower generated by~\eqref{homogeneous}.
		
		Note that the vanishing of $a_R$ or $a_L$ is crucial to the above argument, for if the term proportional to $a_{L}a_R$ is nonzero, then a solution of the form~\eqref{homogeneous} does not exist in the entire domain. In fact, this statement is true even asymptotically: only when~\eqref{StaticLNvanishingcond} is satisfied does the condition $b_{0}=0$ hold. Indeed, if $b_{0}=0$ for a regular solution then, since  $\psi\sim a_0$ is analytic in the \emph{open} interval $(\frac{1}{2},\infty)$, it must be possible to extend this constant solution all the way out to the horizon. Being analytic and bounded in this interval, it must be a constant. But since the last term in~\eqref{scalareq_psi} is not negligible at the horizon unless $a_La_R=0$, this equation implies $\psi_{0}=a_0=0$. Thus, whenever both of these parameters are nonzero, any nontrivial solution must behave asymptotically as
		\begin{equation}\label{generalpsi0}
			\psi_{0}\sim \left(a_0 + \frac{b_0}{\bx}\right)\left[1+\frac{{i}(a_L+a_R)}{\bar{x}} \right],
		\end{equation}
		where both $a_0$ and $b_0$ are nonzero. We can  now use the Ladder operators to raise the solution to an arbitrary level $2n$, and we  find, for any such level, a nontrivial decaying branch with asymptotic behavior $\psi_{2n}^{(2)}\propto b_{0}/\bar{x}^{n+1}$, as can be seen directly by repeated applications of $D_{\ell}^{+}$ on~\eqref{generalpsi0}. We thus see that the conservative tidal response coefficients are nonzero in this case, to the presence of decaying modes at infinity. Similarly, one can use the asymptotic behavior of $\psi_{1}$ to verify that $b_1\neq 0$ for any regular solution, thus giving rise, through the action of $D_{\ell}^{+}$, to a tower of solutions with non-trivial $r^{-\ell-2}$ coefficients at odd $\ell$, hence implying that Love numbers are always nonzero at these levels.  
		
		This analysis hence shows that, not only do conditions~\eqref{StaticLNvanishingcond}
		imply a quasi-polynomial $\Phi_{\ell}$, which is well-known to signal Love number vanishing, but also that these are the \emph{only} conditions consistent with such a behavior. In the following subsections, the response coefficients will be derived in full, and we will see that these conclusions are in fact perfectly consistent with the results obtained through explicit derivation.

		The above discussion implies that static Love number vanishing in this background is equivalent to simply requiring that a homogeneous solution of the Hamiltonian equation exists. Indeed,~\eqref{StaticLNvanishingcond} represent the \emph{only} scenarios in which~\eqref{homogeneous} solves~\eqref{HamiltonianEq} which, as we saw, is a necessary and sufficient condition for the existence of solutions with $b_{\ell}=0$. This offers an alternative explanation for Love number vanishing in the theory, one that is much simpler, though less general, than the Love symmetry analysis that will be developed in Sec.~\ref{sec:LoveSymmetry}. Moreover, using a linear transformation to map the solutions of ~\eqref{HamiltonianEq} into those of an equation of the form $\left[\frac{d^2}{d\bx^2} + V(\bx)\right]\chi_{\ell}(\bx)=0$, we can easily show that the ratio $b_{\ell}/a_{\ell}$ is real, so that the imaginary part that would correspond to the dissipative coefficients vanish, in agreement with the results of the following subsection. As we discuss in further detail in Sec.~\ref{sec:dynamical}, the black hole does in fact possess static dissipative responses, which can be explained by source-response ambiguity. Using the ladder operators and~\eqref{asymptsol}, we find that the imaginary part of the response is generally nonzero when corrections of order $\bx^{\ell-1}$ and $\bx^{-\left(\ell+2\right)/2}$ are respectively added to the growing and decaying branches.

		The ladder structure derived above is related to a hidden symmetry which gives rise to a conserved current and a ladder operator algebra. Indeed, a lengthy but otherwise straightforward calculation shows that the Ladder operators we have derived satisfy the conditions
		\begin{align}
			D_{\ell}^{+} D_{\ell+2}^{-} - K_{\ell} = D_{\ell+4}^{-} D_{\ell+2}^{+} - K_{\ell+2} \hspace{30pt} \text{and} &&  	D_{\ell-2}^{+} D_{\ell}^{-} - K_{\ell-2} = D_{\ell+2}^{-} D_{\ell}^{+} - K_{\ell},
		\end{align}
		where $ K_{\ell}\equiv \left(\frac{\xi_{\ell}}{2}\right)^2+a_L^2+a_R^2+\left(\frac{2a_La_R}{\xi_{{\ell}}}\right)^2$. Using these identities , we can derive the following intertwining relations between ladder operators and Hamiltonians at different levels:
		\begin{align}\label{LadderAlgebra}
			\mathcal{H}_{\ell+2}D_{\ell}^{+}=D_{\ell}^{+}\mathcal{H}_{\ell}, && \text{and}\hspace{50pt}\mathcal{H}_{\ell-2} D_{\ell}^{-} = D_{\ell}^{-} \mathcal{H}_{\ell},
		\end{align}
		which define the algebraic structure associated to the ladder symmetry. These relations can be used to derive, at each level $\ell$, a conserved current. In the special case $a_R=0$ $\ell=2n$ even, the conserved quantity is $\left.J_{2n}(\bx)\right|_{a_R=0}=	\left(\bar{\Delta}\partial_{\bar{x}} + 2{i}a_{L}\right)D_{2}^{-}\dots D_{2n}^{-}\psi_{\ell},$ with an obvious analogue in the $a_L=0$ case. This current has the exact same form as that found for Kerr black holes (see, for example,~\cite{Hui, HuiII, Ghosh}), and can be derived by the same method. 
		
		If $a_L$ and $a_R$ are both nonzero, the quantity discussed above is not a conserved current. We must thus generalize these expressions including new terms to allow for on-shell conservation. To derive the current for even $\ell$ case, we first note that $\mathcal{H}_{0}\psi_{0}=0$ implies
		\begin{equation}
			\partial_{\bar{x}} \left[ \bar{\Delta} \partial_{\bar{x}} \psi_0(\bx) + 2\mathrm{i}(a_L + a_R) \psi_0 +  4a_L a_R \int_{\bar{x}_0}^{\bar{x}} \frac{\psi_0(\bar{x}')}{\bar{x}' + \frac{1}{2}}  \mathrm{d}\bar{x}' \right] = 0,
		\end{equation}
		whence we read off the conserved current at level $\ell=0$
		\begin{equation}
			J_{0}= \left[\bar{\Delta} \partial_{\bar{x}} + 2i \left(a_{L} + a_{R}\right) \right]\psi_{0}(\bx) + 4a_{L}a_{R} \int_{\bar{x}_{0}}^{\bar{x}} \frac{ \psi_{0}(\bx') }{ \bar{x}' + \frac{1}{2} }  \mathrm{d}\bar{x}'.
		\end{equation}
		
		By combining these results with ladder operators, one can derive the general current for even $\ell=2n$ at higher levels:
		\begin{equation}\label{J2n}
			{J_{2n}}=	\left(\bar{\Delta}\partial_{\bar{x}} + 2{i}\left(a_{L} + a_{R}\right)\right)D_{2}^{-}\dots D_{2n}^{-}\psi_{2n} + 4a_{L}a_{R}\int_{\bar{x}_{0}}^{\bar{x}} \frac{D_{2}^{-}\dots D_{2n}^{-}\psi_{2n}(\bar{x}')}{\bar{x}' + \frac{1}{2}} d\bar{x}'.
		\end{equation}

		Since the lowering operators we derived always shift $\ell$ by two, solutions with odd $\ell$ do not belong to the same tower as $\psi_{0}$, and the above current therefore does not apply to them. It is, however, simple to perform an analogous derivation starting from $\psi_{1}(\bx)$, in which case we find
		\begin{equation}\label{J2n1}
			\begin{split}
				J_{2n+1} &= \left( \bar{\Delta} \partial_{\bar{x}} + 2\mathrm{i}\left(a_{L} + a_{R}\right) \right) \left( D_{3}^{-}\cdots D_{2n+1}^{-} \psi_{2n+1} \right) + \\ & \ \ \ \ \ \ \int_{\bar{x}_{0}}^{\bar{x}}  \left( \frac{4a_{L}a_{R}}{\bar{x}' + \frac{1}{2}} - \frac{3}{4} \right)\left( D_{3}^{-}\cdots D_{2n+1}^{-} \psi_{2n+1}(\bx')\right)\mathrm{d}\bar{x}'
			\end{split}
		\end{equation}
		
		\cmmnt{
			\begin{equation}
				\partial_{\bar{x}} J_{1} = \underbrace{\partial_{\bar{x}} \left[ \bar{\Delta} \partial_{\bar{x}} \psi_1 + 2\mathrm{i}(a_L + a_R) \psi_1 \right]}_{-\left( \frac{4a_L a_R}{\bar{x} + \frac{1}{2}} - \frac{3}{4} \right) \psi_1} + \underbrace{\partial_{\bar{x}} \left[ \int_{\bar{x}_0}^{\bar{x}} \left( \frac{4a_L a_R}{\bar{x}' + \frac{1}{2}} - \frac{3}{4} \right) \psi_1(\bar{x}')  \mathrm{d}\bar{x}' \right]}_{\left( \frac{4a_L a_R}{\bar{x} + \frac{1}{2}} - \frac{3}{4} \right) \psi_1} = 0,
			\end{equation}
			hence
			\begin{equation}\
				J_{2n+1} = \left[\left( \bar{\Delta} \partial_{\bar{x}} + 2\mathrm{i}\left(a_{L} + a_{R}\right) \right)+ \\ \int_{\bar{x}_{0}}^{\bar{x}}d\bx' \left( \frac{4a_{L}a_{R}}{\bar{x}' + \frac{1}{2}} - \frac{3}{4} \right)  \right]D_{3}^{-}\cdots D_{2n+1}^{-} \psi_{2n+1}(\bar{x}')
			\end{equation}
		}
		\noindent as the conserved current at odd levels. These results highlight the fact that even and odd values of $\ell$ belong to separate towers of solutions in this framework, a distinction that did not exist in the usual four-dimensional black holes. Note that, unlike what is found for $a_{R/L}=0$, these currents are manifestly nonlocal due to the integral in the last term. Nevertheless, these are still valid currents, and the integrals can be addressed with appropriate techniques. One simple way to obtain the currents explicitly is to analyze the solution in an arbitrarily small neighborhood defined by $\frac{1}{2}<\bx<\delta$ and take $\bar{x}_{0}=1/2$. Since we already know $\psi_{\ell}$ in closed-form, this is a simple task in our case. Choosing $\delta$  arbitrarily small, one can identify the solution with the known constant value $\psi_{0}(0)=1$. Since no singularities exist in the interval $(\frac{1}{2},\delta)$, the integral in~\eqref{J2n} vanishes due to the vanishing domain of integration, and the current is easily evaluated to $J_{0}=2i\left(a_L+a_R\right)$. To obtain the current at higher levels, we need only apply the ladder operators recursively to lower $\psi_{2n}$ all the way to $0$ and thus obtain 
		\begin{equation}\label{Current_even}
			J_{2n} = 2i(a_L + a_R) \prod_{k=1}^{n} \left( i(a_L + a_R) + \frac{2a_L a_R}{k} - \frac{k}{2} \right),
		\end{equation}
		valid for $n>0$. In the odd case, we carry out an identical procedure to find $J_{1}=2i\left(a_L+a_R\right)$, which is interestingly exactly the same ``seed" current as in the even case. At higher levels, we find
		\begin{equation}
			J_{2n+1} =2i(a_L + a_R) \prod_{k=1}^{n} \left( i(a_L + a_R) + \frac{4a_L a_R}{2k+1} - \frac{2k+1}{4} \right),
		\end{equation}
		which is again valid for $n> 0$. The above conserved currents are useful, as they may allow one to extract information of the solution, including its asymptotic or near-horizon behavior, without explicitly solving the full equation.  In this way, these currents  may be  useful tools in the process of matching the behavior of solutions from different regions of spacetime. This reasoning may at first appear cyclical because we have used the exact solutions of the scalar equation to derive the Ladder operators. In fact, the explicit Hamiltonian, which is known from the wave equation, and the intertwining relations~\eqref{LadderAlgebra} suffice to derive both the ladder operators and currents, up to irrelevant constant factors. Thus, we could just as easily have solved these algebraic equations and imposed consistency with the scalar equation to derive all the main results of this section. This observation is important, because it allows for the extension of the ladder structure considered here to other models (or, in the near-zone generalization, different splits within the same theory) where a closed-form solution cannot be found. 
		
		Another interesting property of four-dimensional black holes is the relationship between the conserved currents and the dissipative coefficients of the black hole. In fact, these quantities can be shown to be equivalent, differing at most by a proportionality constant depending on the adopted conventions~\cite{Hui, HuiII}. In $5D$ black holes, the situation seems to be more complicated outside of the cases covered by conditions~\eqref{StaticLNvanishingcond}, which of course give the same results as Kerr. When $a_La_R\neq 0$, we can use the static response coefficients~\eqref{klstatic} derived in Sec.~\ref{sec:dynamical} to formally write $k_{\ell}=\frac{\Gamma(-1-\ell)}{i(a_L+a_R)}F(a_R, a_L,\ell)J_{\ell}$, where $F(a_R, a_L, \ell)$ is a nonzero and nondivergent coefficient, with the pole from the Gamma function being a consequence of the divergence of static Love numbers in $5D$ when $a_La_R\neq 0$. If this relationship were known \emph{a priori}, even without explicit deduction of  $F(a_R, a_L \ell)$, it would in principle be possible to deduce the general behavior of the intrinsic response coefficients. The pole from $\Gamma(-1-\ell)$ can only be compensated when the $a_La_R$ is zero and $\ell$ is even, which leads us back to conditions~\eqref{StaticLNvanishingcond}. On the other hand, the above relationship implies that $\rm{Im}(k_{\ell})=0$ if and only if $J_{\ell}$ vanishes, while the real part of $k_{\ell}$ remains divergent whenever $a_La_R\neq 0$, regardless of the value of $J_{\ell}$ . These results are consistent with those we shall derive explicitly in Sec.~\ref{sec:dynamical} through analytic continuation, where it will be seen that static dissipative responses vanish only when $a_L+a_R=0$, precisely the same condition that defines the the zeroes of $J_{\ell}$. This relationship merits further investigation in other systems to decide whether it is a general feature among reasonable 5D black holes. If it is, then the currents could in principle be used to infer some of the general features of the tidal coefficients, such as the zeroes of its dissipative component. At the very least, the above discussion seems to suggest some connection between the ladder structure of the static equation and the intrinsic black hole responses, despite the fact that analytical continuation has not been used in the results of this section.

		\cmmnt{
			Indeed, taking the complex conjugate of~\eqref{HamiltonianEq} can construct the bilinear equation
			\begin{equation}
				\psi_{\ell}^*\mathcal{H}_{\ell}\psi_{\ell}-	\psi_{\ell}(\mathcal {H}_{\ell}\psi_{\ell})^*=0,
			\end{equation}
			which after some algebra leads to a conservation equation for the current
			\begin{equation}
				j_\ell(\bar{x}) = \bd \left( \psi_\ell^* \partial_{\bx} \psi_\ell - \psi_\ell \partial_{\bx} \psi_\ell^* \right) + 2{i}(a_L+a_R)|\psi_\ell|^2,
			\end{equation}
			which  may be simpler to use in applications  due to the fact that it only depends on the wavefunctions and their derivatives. }

		The Ladder formalism we have derived has allowed us to deduce the conditions of Love number vanishing, leading in a simple way to many of the general conclusions that we will derive afterwards with use of more sophisticated techniques. Moreover, we have also derived, at each level $\ell$, a conserved current whose on-shell conservation follows from the Hamiltonian equation~\eqref{HamiltonianEq}, and which could in principle be used to infer important information about the black hole's tidal perturbations without recourse to the full solution. In general, this shows that the raising operators provide useful tools that can enhance our understanding of black hole tidal perturbations through the use of relatively simple techniques. By matching the behavior at different regions, we could in principle derive the asymptotic coefficients explicitly and try to deduce the tidal response coefficients, although that may not be a simple task. We shall however find these parameters in the following subsection by direct evaluation of the asymptotic form of the solutions.

		\cmmnt{and
			\be
			D^-_{\ell+2}D^+_{\ell}=D^+_{\ell-2}D^-_{\ell} +\frac{1}{4}(2\xi_{\ell}-1)\left[1+\frac{a_L^2a_R^2}{\xi_{\ell}^2(\xi_{\ell}-1)^2}\right]\,.
			\ee}

		\subsection{\bf Tidal response coefficients in the static case}
		
		The results obtained above can be used to derive the tidal response coefficients in the static case ($\om=0$). As shown, the scalar solution~\eqref{eq:phiin0} is written in terms of a hypergeometric function with parameters $a=\xi_{\ell}+2i a_R$, $b=\xi_{\ell}+2i a_L$ and $c=2i(a_R+a_L)+1$, so that $a+b-c=2\xi_{\ell}-1=\ell+1\in\mathbb{N}$. Let us first consider the case where one of the parameters $a_{L/R}$ vanishes and $\ell=2n$ is an even natural number. In this case, $a$ becomes an integer, and it follows that the hypergeometric function we are dealing with becomes a polynomial of degree $n$~\cite{highertranscendental}. This can be verified with use of the transformation formula
		\begin{equation}
			{}_2F_1(a, b; c; z) = (1 - z)^{c - a - b}  {}_2F_1(c - a, c - b; c; z),
		\end{equation}
		which leads to
		\begin{equation}
			\psi_{2n}(z) = \left( 1-z \right)^{-n} 
			{}_{2}F_{1}\left( 
			-n + 2i\bar{a}_{L},\ 
			-n;\ 
			1 + 2i\bar{a}_{L};z
			\right).
		\end{equation}
		
		In this form, it is seen that the second coefficient is a negative integer, so that the hypergeometric series terminates after the $n+1$ terms. The radial solutions thus display the quasi-polynomial behavior that follows the vanishing of conservative responses in this case, in agreement with our previous results. In terms of $\bx$, these solutions are written:
		\begin{equation}\label{quasipolynomial}
			\Phi_{2n}(\bar{x}) = \mathcal{A}_{2n} \sum_{k=0}^{n} \frac{(2ia_L - n)_k (-n)_k}{(1+2ia_L)_k \, k!} \left(\bar{x}-\frac{1}{2}\right)^{k+ia_L} \left(\bar{x}+\frac{1}{2}\right)^{n-k-ia_L}
		\end{equation}
		which at large $\bx$ is dominated by the $\bx^{n}$ term, thus confirming the expected $r^{\ell}$ behavior at infinity.

		Next, let $a_L,a_R\neq 0$. For natural $a+b-c$ and $|z|<1$, the hypergeometric function can be written as the series \cite{Abramowitz}:
		\begin{equation}
			\begin{split}
				&{_{2}F_{1}}(a,b,a+b-n,z)={\G(n)\G(a+b-n)\over \G(a)\G(b)}(1-z)^{-n}\sum^{n-1}_{j=0}{(a-n)_{j}(b-n)_{j}\over j!(1-n)_j}(1-z)^j\\ & -{(-1)^n\G(a+b-n)\over \G(a-n)\G(b-n)}\sum^{+\infty}_{j=0}{(a)_{j}(b)_{j}\over j!(n+j)!}(1-z)^{j}\left[\ln(1-z)  -\psi(j+1)-\psi(j+n+1)\right. \\& \left. \hspace{120pt} +\psi(a+n)+\psi(b+n)\right],\label{HG_abn}
			\end{split}
		\end{equation}
		where $\G(y)$ is the Gamma-function of $y$, $(y)_{j}={\G(y+j)/ \G(y)}$ is the Pochhammer symbol and $\psi(y)={\G'(y)/\G(y)}$ is the digamma-function  and $n=\ell+1$. In order to extract the response coefficients of the Black hole, we must consider the behavior of the perturbation at infinity, which in this sense we identify with the location of the source of the external tidal potential. Taking the leading-order terms in the asymptotic expansion implied by (\ref{HG_abn}), we obtain
		\begin{multline}
			{_2F_{1}}(a,b,a+b-n,z) \underset{z\to 1}{\longrightarrow} \quad{\G(n)\G(a+b-n)\over \G(a)\G(b)}(1-z)^{-n}-{(-1)^n\G(a+b-n)\over n!\G(a-n)\G(b-n)}\ln{(1-z)}.
		\end{multline}
		Using this relation we easily write the asymptotic behavior of the radial wave function $\Phi_{\ell}(r)$, which is of pivotal to derive the response coefficients. This behavior is as follows:
		\begin{multline}\label{AsymptoticRl}
			{\Phi}_{\ell}(r)\underset{r\to\infty}{\sim}{\G(\ell+1)\G(a+b-\ell-1)\over\G(a)\G(b)}\vk^{-\ell/2}r^{\ell}\left(1+(-1)^{\ell}\times\right.\\\left.{\G(a)\G(b)\over \ell!(\ell+1)!\G(a-\ell-1)\G(b-\ell-1)}\vk^{\ell+1}\ln\left({\vk\over r^2-r^2_{-}}\right)r^{-2(\ell+1)}\right),
		\end{multline}
		where we have defined
		\begin{equation}
			\vk\equiv r^2_{+}-r^2_{-}
		\end{equation}
		
		Following the usual approach, we find the response coefficients by writing the asymptotic series $\Phi_{\ell}$ and then evaluating the ratio between the coefficients from the  $\sim r^{\ell}$ and $\sim r^{-\ell-2}$ terms. This ratio can be directly  extracted from the second row of~\eqref{AsymptoticRl}, whence we obtain
		\be 
		\lambda_{\ell}(\om=0)=(-1)^{\ell}{\G\left(1+{\ell\over 2}+2ia_{R}\right)\G\left(1+{\ell\over 2}+2ia_{L}\right)\over \ell!(\ell+1)!\G\left(-{\ell\over 2}+2ia_{R}\right)\G\left(-{\ell\over 2}+2ia_{L}\right)}\vk^{\ell+1}\ln\left({\vk\over r^2-r^2_{-}}\right).\label{Love_n_st}
		\ee
		One important property of the above equation is the fact that, due to relation~\eqref{eq.ok}, the response coefficients can be described entirely in terms of parameters from the seed Myers-Perry black hole. Moreover, they show the characteristic logarithmic running of black hole tidal response. This behavior is owed to the specific relationship between the hypergeometric function's parameters, namely to the fact that, as already pointed out above, $a+b-c$ is a natural number. As we show in the next section, this condition also holds true for the near-zone equation.
		
		The imaginary part of $\lambda_{\ell}$ is given by
		\begin{equation}\label{ImLambda}
			\frac{i \pi ^2 C_{\ell}}{2 \ell! \Gamma (\ell+2) \Gamma \left(-2 i a_L-\frac{\ell}{2}\right) \Gamma \left(2
				i a_L-\frac{\ell}{2}\right) \Gamma \left(-2 i a_R-\frac{\ell}{2}\right) \Gamma \left(2 i a_R-\frac{\ell}{2}\right)},
		\end{equation}
		where
		\begin{equation}
			\begin{split}
				C_{\ell}\equiv &\frac{2}{\cosh (2 \pi  (a_L-a_R))+(-1)^{\ell+1} \cosh (2 \pi 
					(a_L+a_R))} \\ &+\text{csch}\left(2 \pi  a_L-\frac{i \pi  \ell}{2}\right) \text{csch}\left(2 \pi 
				a_R-\frac{i \pi  \ell}{2}\right)
			\end{split}.
		\end{equation}
		
		Since we are dealing with the cases $a_L,a_R\neq 0$, all Gamma functions in the denominator of~\eqref{ImLambda} are finite and nonzero, so that the vanishing of the dissipative response must follow from that of $C_{\ell}$ for all $\ell\in\mathbb{N}$. To verify that this is indeed the case, let us consider the even and odd cases separately. Letting $\ell=2n$, we can use the elementary identities $\cosh(x-y)-\cosh(x+ y)=-2\sinh(x)\sinh(y)$ and $\sinh(x -in\pi)=\sinh(x)\cosh(-n\pi)=(-1)^n\sinh(x)$ for integer $n$ to write:
		\begin{equation}
			C_{2n}=\frac{2}{-2\sinh(2\pi(a_R))\sinh(2\pi(a_L))} + \frac{1}{(-1)^n\sinh(2\pi(a_R))\sinh(2\pi(a_L))(-1)^k}=0. 
		\end{equation}
		On the other hand, if $\ell=2n+1$, we use  $\cosh(x-y)-\cosh(x+ y)=2\cosh(x)\cosh(y)$ and $\sinh(x-i\pi(n +1/2))=-i(-1)^n\cosh(x)$ to write:
		\begin{equation}
			C_{2n+1}=\frac{1}{\cosh(2\pi(a_R))\cosh(2\pi(a_L))} + \frac{1}{(-i)^2(-1)^{2n}\cosh(2\pi(a_R))\cosh(2\pi(a_L))}=0,
		\end{equation}
		so that the dissipative coefficients are zero for all physical solutions of the static Klein-Gordon equation in the STU background we are dealing with.
		
		In the limit $\delta_1=\delta_2=\delta_3\to0$, the above results are in good agreement with those found in~\cite{PRD108}, and generalize the tidal coefficients of that work to the case of nonzero charges. On the other hand, the response coefficients found in Ref.~\cite{Charalam_2023} for Myers-Perry black holes are not the same, as the authors of that paper have identified a nonzero dissipative response even in the static case, as well as divergent Love numbers that do not agree with the $\lambda_{\ell}$ found in this section. Rather than a contradiction, we argue this difference is in fact a result of the two different methods employed, and while the conservative part of both quantities have been referred to as ``Love numbers" throughout the literature, these two coefficients have different physical meaning. Namely, the authors in Ref.~\cite{Charalam2024} have used analytic continuation to derive response coefficients in such a way as to avoid the ambiguity that emerges due to overlapping in the power series expansions of relative to the source and response contributions to $\Phi_{\ell}$. The so-derived coefficients represent the inherent black hole response and give rise to the ``canonical" Love numbers, while the coefficients~\eqref{Love_n_st}  may be identified with the total physical response functions. We shall discuss these issues more thoroughly in the following section, where we will derive the response coefficients using both the method employed above and that of analytic continuation approach, with the static perturbations included as the case $\omega=0$. To avoid possible ambiguity, we shall, both in the static and dynamical cases, use the symbol $\lambda_{\ell}$ for the total response coefficients, obtained through the method carried out in this subsection, while for the ``intrinsic" or canonical coefficients the symbol $k_{\ell}$ will be used.
		
		\section{Dynamical Love numbers and Love symmetry in the near-zone regime}\label{sec:dynamical}
		
		\subsection{Love numbers in near-zone approximation}
		
		The results from the previous section were valid for the static case, which amounts to time-independent perturbations. We can also extend the investigation to the dynamical scenario by considering the so-called near-zone limit. This regime corresponds to long-wavelength perturbations within a finite distance from the horizons. More precisely, we require the simultaneous fulfillment of the conditions:
		\begin{subequations}\label{NearZoneDef}
			\begin{align}
				\omega r_{+} &\ll 1 \label{NearZoneDef1} \\
				\omega\left(r-r_{+}\right) &\ll 1 . \label{NearZoneDef2}
			\end{align}
		\end{subequations}
		
		The first of these sets the scale by precisely defining the ``long-wavelength" condition globally, which is achieved by relating the frequency to the fixed parameter $r_{+}$. The second equation sets the range of validity of the near-zone approximation, which breaks down at sufficiently large distances from the horizon, with the precise meaning of ``large" implied by the first condition in terms of $\omega$. If $r$ is of the same order as $r_{+}$ the first condition implies  $\om x\ll 1$, and we also see that the near-zone approximation is always valid sufficiently close to the horizon.  In the exterior region, we have  a range of validity $\frac{1}{2}\leq\bar{x} \ll \frac{1}{\omega^2 (r_{+}^2 - r_{-}^2)}$. Thus, between the near-horizon and asymptotically flat regimes, there exists a clear overlap region in which both of the assumptions hold. The near-zone equation may thus be valid for relatively large frequencies (as long as $r_{+}$ is sufficiently small for~\eqref{NearZoneDef1} to remain valid) if $\bx$ is small enough, while it may also extend to large distances from the horizon if sufficiently small frequencies are considered. 
		\cmmnt{\begin{equation}\label{overlapregion}
				\frac{1}{2}\leq\bar{x} \ll \frac{1}{\omega^2 (r_{+}^2 - r_{-}^2)}
		\end{equation}}
		
		The near-zone equation is found by defining a \emph{split}, which requires writing the wave equation operator of the Klein-Gordon equation in the form $\mathcal{\hat D_{K.G}}\equiv\mathcal{\hat D}_0+\mathcal{\hat D}_1$. We require identity between $\left.\mathcal{\hat D}_0\right|_{\omega=0}$ and the scalar wave operator, while $\mathcal{\hat D}_1$ should amount to contributions that can be neglected within the range of validity of the near-zone regime. The desired approximation is achieved by taking $\mathcal{\hat D_{K.G}}\approx\mathcal{\hat D}_0$, i.e., by the neglecting of $\mathcal{\hat D}_1$. As is well-known, this split is not uniquely defined by these requirements, and different splits should amount to distinct regimes of validity, although these different choices must agree as $\omega\to 0$ and within a small neighborhood of the horizon.
		
		To define our near-zone split, let us write 
		\begin{equation}
			\hat{\Lambda}^{(\scalebox{1}{$\epsilon$})}= \vec{K}^2 + \epsilon\omega^2 \left[(l^2_1+l_2^2)+(l^2_2-l_1^2)\cos 2\theta\right]
		\end{equation}
		in the angular equation and

		\begin{equation}
			V^{(\scalebox{1}{$\epsilon$})}(\bx)=V_0(\bx) + \epsilon V_1(\bx)
		\end{equation}
		in the radial equation, where 
		
		\begin{equation}
			V_0(\bx)=\frac{\left(\om-2(m_{R}\Om^{R}_{+}+m_{L}\Om^L_{+})\right)^2}{4\k^2_{+}\left(\bx-\frac{1}{2}\right)}-\frac{\left(\om-2(m_{R}\Om^{R}_{-}+m_{L}\Om^L_{-})\right)^2}{4\k^2_{-}\left(\bx+\frac{1}{2}\right)}-\frac{\Lambda_{l}}{4}
		\end{equation}
		and
		\begin{equation}
			V_1(\bx)=\frac{1}{4}\left(\bx\Delta\omega^2+M\omega^2\right) 
		\end{equation}
		so that the full scalar equation is achieved by setting $\epsilon=1$, while we define the near-zone regime as the limit $\epsilon\to0$. For the angular equation, this amounts to the identification between $\hat{\Lambda}$ and the $3$-sphere Laplace operator, while the radial split neglects the $\bx$-independent terms proportional to $\omega^2$ and $\bar{x}\omega^2$ which must both be small in the near-zone region. 
		
		In the near-zone approximation, the radial equation is reduced to:
		\cmmnt{
			\begin{equation}
				\frac{\p}{\p\bx}\left(\bx^2-\frac{1}{4}\right)\frac{\p {\Phi}_{l}(\bx)}{\p\bx}+\left(\frac{\left(\om-2(m_{R}\Om^{R}_{+}+m_{L}\Om^L_{+})\right)^2}{4\k^2_{+}\left(\bx-\frac{1}{2}\right)}-\frac{\left(\om-2(m_{R}\Om^{R}_{-}+m_{L}\Om^L_{-})\right)^2}{4\k^2_{-}\left(\bx+\frac{1}{2}\right)}-\frac{\Lambda_{l}}{4}\right){\Phi}_{l}(\bx)=0.
		\end{equation}}
		\begin{equation}
			\left[\pb{\bx}\left(\bd\pb{\bx}\right)+\left(\frac{\left(\om-2(m_{R}\Om^{R}_{+}+m_{L}\Om^L_{+})\right)^2}{4\k^2_{+}\left(\bx-\frac{1}{2}\right)}-\frac{\left(\om-2(m_{R}\Om^{R}_{-}+m_{L}\Om^L_{-})\right)^2}{4\k^2_{-}\left(\bx+\frac{1}{2}\right)}-\frac{\Lambda_{l}}{4}\right)\right]{\Phi}_{l}(\bx)=0.\label{eq.nzeq}
		\end{equation}
		
		We note that~\eqref{eq.nzeq} has exactly the same form as the static equation~\eqref{eq.str} investigated in the previous section, as the generalization to nonzero $\omega$ amounts to shifts in the coefficients of the hypergeometric equation above, which do not fundamentally alter its structure as a differential equation in $\bar{x}$. This means that all results from the previous section, including the ladder formalism, can be generalized to the near-zone regime by simply taking these shifts into account. We also point out that, since the near-zone angular equation is independent of $\omega$, the constant $\Lambda_{l}$ in the equation (\ref{eq.nzeq})  is the same as in the static case. 
		
		The above equation can be solved in the same way as its static analogue, giving rise to a solution of the form:
		\begin{multline}
			{\Phi}_{\ell}(\bx)=B_{\ell}\left(\bx-{1\over 2}\over \bx+{1 \over 2}\right)^{-i{\om\over 2\k_{+}}+i(a_{R}+a_{L})}\left(\bx+{1\over 2}\right)^{-\xi_{\ell}}\times\\{_2F_{1}}\left(\xi_{\ell}-i{\om\over 2}p+2ia_{R}, \xi_{\ell}-i{\om\over 2}q+2ia_{L};1-i{\om\over\k_{+}}+2i(a_R+a_L);{\bx-{1\over 2}\over \bx+{1\over 2}}\right).\label{solut_nz}
		\end{multline}
		where  $\xi_{\ell}$, $a_{R}$ and $a_{L}$ are as defined in the previous section, and we have introduced the parameters
	\begin{subequations}\label{Def:p_and_q}
	\begin{align}
	&p\equiv{1\over\k_+}+{1\over\k_-}=	\dfrac{\mu\left(\Pi_{\mathrm{c}} + \Pi_{\mathrm{s}}\right)}{\sqrt{\mu - \left(l_{1} + l_{2}\right)^{2}}} \label{Def:p} , \\
	&q\equiv\frac{1}{\k_+}-\frac{1}{\k_-}=\dfrac{\mu\left(\Pi_{\mathrm{c}} - \Pi_{\mathrm{s}}\right)}{\sqrt{\mu - \left(l_{1} - l_{2}\right)^{2}}}
	\label{Def:q},
\end{align}
	\end{subequations}
	where \(\Pi_{\mathrm{c}} \equiv \prod\limits_{i=1}^{3} \cosh\delta_i\) and \(\Pi_{\mathrm{s}} \equiv \prod\limits_{i=1}^{3} \sinh\delta_i\). In the static limit, defined by $\om=0$, the solutions (\ref{solut_nz}) are reduced to (\ref{eq:phiin0}). We note that (\ref{solut_nz}) satisfies the proper boundary conditions at both the horizon, where the radial function should be an ingoing wave, and at infinity. As we mentioned in the previous section, there are two paths we can follow in order to obtain the response coefficients from an asymptotic expansion. Let us first proceed in the same manner as in the previous section and obtain the response coefficients directly from the physical wavefunction i.e., treating $\ell$ as an integer from the start. The parameters of $_2F_1(\ma,\mb;\mc;z)$ in the near-zone regime are $\ma$, $\mb$ and $\mc$ and their counterparts $a$, $b$ and $c$ in the static case are related by:
		\be\label{shifts}
		\ma=a-i{\om\over 2}p,\quad \mb=b-i{\om\over 2}q,
		\quad \mc=c-i{\om\over\k_{+}}.
		\ee
		which give rise to the same relation $\ma+\mb-\mc=a+b-c=l+1$. 
		Since this combination is a nonnegative integer, the analysis of the solution (\ref{solut_nz}) does not really differ from the static case. We are thus able to write the asymptotic expression for the radial wave-function ${\Phi}_{\ell}(r)$ in the form:
		\begin{multline}
			{\Phi}_{\ell}(r)\underset{r\to\infty}{\sim}{\G(\ell+1)\G(\ma+\mb-\ell-1)\over\G(\ma)\G(\mb)}\vk^{-\ell/2}r^{\ell}\left(1+(-1)^{\ell}\times\right.\\\left.{\G(\ma)\G(\mb)\over \ell!(\ell+1)!\G(\ma-\ell-1)\G(\mb-\ell-1)}\vk^{\ell+1}\ln\left({\vk\over r^2-r^2_{-}}\right)r^{-2(\ell+1)}\right),
		\end{multline}
		
		From the above asymptotic behavior, we derive the {response coefficients}, which can be written in the form:
		\be 
		\lambda_{\ell}(\om)=(-1)^{\ell}{\G\left(1+{\ell\over 2}-i{\om\over 2}p+2ia_{R}\right)\G\left(1+{\ell\over 2}-i{\om\over 2}q+2ia_{L}\right)\over \ell!(\ell+1)!\G\left(-{\ell\over 2}-i{\om\over 2}p+2ia_{R}\right)\G\left(-{\ell\over 2}-i{\om\over 2}q+2ia_{L}\right)}\vk^{\ell+1}\ln\left({\vk\over r^2-r^2_{-}}\right)\label{Love_n},
		\ee
		where, as before, $a_L$ and $a_R$ are given by~\eqref{eq.aRaL}. The above expression provides a natural generalization of the static coefficients (\ref{Love_n_st}), and we note that the relation (\ref{Love_n}) is valid as long as the conditions that define the  near-zone approximation are met with sufficient precision. The form (\ref{Love_n}) is not convenient for practical purposes, and we can rewrite these coefficients in a more tractable form with the use of mirror formulae and other Gamma-function identities. We also point out that final explicit relation for the Love numbers depend on the parity of $\ell$. If $\ell=2n$, where $n\in\mathbb{N}$, we can use the Gamma-function identities appropriate to the present case \cite{Abramowitz, Perry_arx23} to write
		\be
		{\G\left({\ell\over 2}+1+iA\right)\over \G\left(-{\ell\over 2}+iA\right)}={\G\left(n+1+iA\right)\over \G\left(-n+iA\right)}=i(-1)^n\pi A\prod^{n}_{k=1}(k^2+A^2).\label{gamma_rel_even}
		\ee
		
		Now it will prove useful to introduce the variables
		\begin{equation}
			y_1 = 2a_R - \frac{\omega}{2}p \qquad \text{and} \qquad y_2 = 2a_L - \frac{\omega}{2}q,
		\end{equation}
		which together account for the totality of the frequency and boost dependence on the tidal response coefficients. Using~\eqref{gamma_rel_even} and these variables, we can write the even $\ell$ coefficients in the form:
		\begin{equation}
			\lambda_{2n}=-\frac{\pi^2 y_1 y_2}{(2n)!(2n+1)!}\varkappa^{2n+1}\left[\prod^{n}_{j=1}\left(j^2+y_1^2\right)\left(j^2+y_2^2\right)\right]\ln{\left(\frac{\varkappa}{r^2-r^2_{-}}\right)}\label{Love_even}
		\end{equation}
		\cmmnt{\begin{multline}
				\lambda_{2n}(\om)=-\pi^2{\left(2a_{R}-{\om\over 2}p\right)\left(2a_{L}-{\om\over 2}q\right)\over (2n)!(2n+1)!}\vk^{2n+1}\\  \times\left[\prod^{n}_{j=1}\left(j^2+\left(2a_{R}-{\om\over 2}p\right)^2\right)\left(j^2+\left(2a_{L}-{\om\over 2}q\right)^2\right)\right]\ln{\left({\vk\over r^2-r^2_{-}}\right)}\hspace{50pt}\label{Lov_even}
		\end{multline}}

		For $\ell=2n+1$ instead of relation (\ref{gamma_rel_even}), we have:
		\be
		{\G\left({\ell\over 2}+1+iA\right)\over \G\left(-{\ell\over 2}+iA\right)}={\G\left(n+1+{1\over 2}+iA\right)\over \G\left(-n-{1\over 2}+iA\right)}=(-1)^{n+1} \prod^{n+1}_{k=1}\left(\left(k-{1\over 2}\right)^2+A^2\right).\label{gamma_rel_odd}
		\ee
		Thus, for odd $\ell$, the \textcolor{black}{response coefficients} are
		\begin{equation}
			\lambda_{2n+1}=-\frac{\varkappa^{2(n+1)}}{(2n+1)!(2n+2)!}\left[\prod^{n+1}_{j=1}\left(\left(j-\frac{1}{2}\right)^2+y_1^2\right)\left(\left(j-\frac{1}{2}\right)^2+y_2^2\right)\right]\ln{\left(\frac{\varkappa}{r^2-r^2_{-}}\right)}
		\end{equation}
		\cmmnt{\begin{multline}
				\lambda_{2n+1}(\om)=-{\vk^{2(n+1)}\over (2n+1)!(2n+2)!}\ln{\left({\vk\over r^2-r^2_{-}}\right)}\\ \times\left[\prod^{n+1}_{j=1}\left(\left(j-{1\over 2}\right)^2+\left(2a_{R}-{\om\over 2}p\right)^2\right)\left(\left(j-{1\over 2}\right)^2+\left(2a_{L}-{\om\over 2}q\right)^2\right)\right] \hspace{20pt}\label{Love_odd}
		\end{multline}}

		\cmmnt{The \textcolor{black}{response coefficients} are therefore are real for all integer choices of $\ell$. Thus, the dissipation part is absent, and the above parameters may be identified with the Love numbers. We note that the imaginary part of dynamical \textcolor{black}{tidal response coefficients} for scalar ( i.e., $s=0$) particles are also known to vanish in the case of Kerr black holes \cite{Perry_arx23}, as the effects that would lead to non-conservative tidal response are only encountered at higher spins. }
		
\begin{figure}[h]
			\centering
			\includegraphics[width=8.5cm]{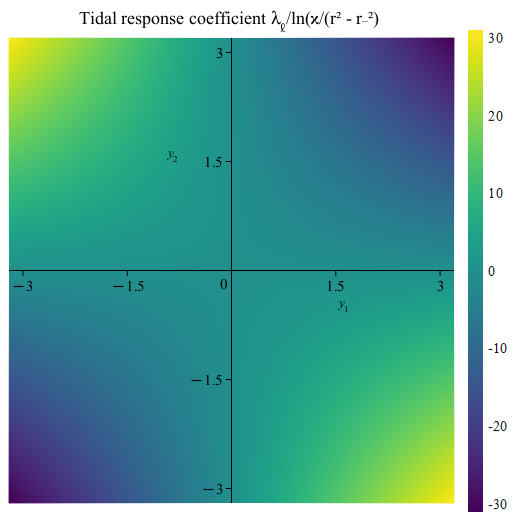}
			\hspace{0.25cm}
			\includegraphics[width=8.5cm]{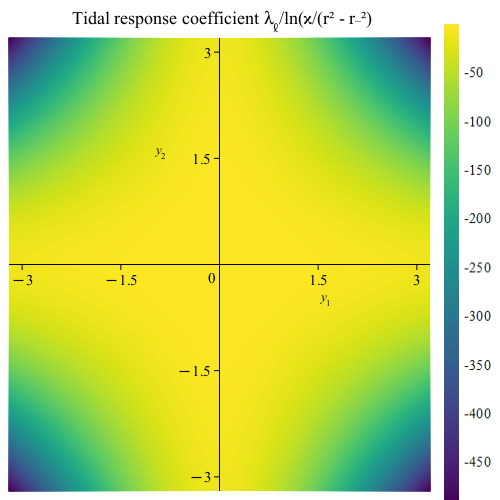}
			\caption{Contour plot representing the real tidal response coefficients~\eqref{Love_n} for the $\ell=0$ (left) and $\ell=1$ (right), as functions of $y_1$ and $y_2$. We present the results divided by the logarithmic factor $\ln(\vk/r^2-r_{-}^2)$, which is independent of $y_1$ and $y_2$, and use the choices $r_{-}=1$, $r_{+}=2$.}
			\label{ResponseCoefficients}
		\end{figure}

		The response coefficients are therefore real for all integer choices of $\ell$, so that the dissipation part is absent from these expressions. We depict these coefficients, normalized by the logarithmic factor which is fixed for each given $r$, in Fig.~\ref{ResponseCoefficients}, where the cases $\ell=0$ and $\ell=1$ are depicted, representing the even and odd cases respectively. Solutions with higher multipole moments present an analogous behavior, namely a ``saddle-like" form for even $\ell$, while the odd coefficients are strictly negative and depend only on the absolute values of $y_{1}$ and $y_{2}$. In both situations the coefficients are seen to increase in magnitude as $|y_{i}|$ increases. Thus, for fixed $\delta_i$, the absolute value of $\lambda_{\ell}$ increases in a manner analogous to what is found in neutral black holes, while for each fixed $\omega$ the magnitude (and, in the case of even $\ell$, the sign) of these coefficients is determined by the interplay between left and right inverse temperatures, which depend on the boosts as seen from definitions~\eqref{eq:betal} and~\eqref{eq:betar}.

		As mentioned in the previous section, the canonical Love numbers can be derived through analytic continuation. This is achieved by treating the multipole index $\ell$ as an arbitrary real number throughout the entire calculation of the response coefficients, and then obtaining the physical responses by taking the limit as $\epsilon \to 0$ of the expression $\ell= N+\epsilon$, with $N\in\mathbb{N}$. This technique has been successful in obtaining Love numbers for various important black holes in the literature, see, for example~\cite{LeTiec, Perry_arx23, De Luca, Barura, Charalam2024,BHstereotyping, Charalam_2023}  and references therein. Through analytic continuation, we are able to address the source-response ambiguity by avoiding the degeneracy that would normally arise from integer coefficients through the Frobenius method, as the overlap between the two series does not arise unless $\ell\in\mathbb{Z}$. 
		
		If the multipole moments are treated as real numbers, we can use the transformation formula
		\begin{equation}
			\begin{split}
				_2F_1(a,b;c;z) = &\frac{\Gamma(c)\Gamma(c-a-b)}{\Gamma(c-a)\Gamma(c-b)}\,_2F_1(a,b;a+b-c+1;1-z) \\
				&+ (1-z)^{c-a-b}\frac{\Gamma(c)\Gamma(a+b-c)}{\Gamma(a)\Gamma(b)}\,_2F_1(c-a,c-b;c-a-b+1;1-z),
			\end{split}
		\end{equation}
		which, after expansion at infinity, gives the response coefficients
		\begin{equation}\label{kl}
			k_{\ell} = \vk^{1+\ell} \frac{\Gamma(-1-\ell)}{\Gamma(1+\ell)} \frac{\Gamma\left(1+\frac{\ell}{2} - i\frac{\omega}{2}p + 2ia_R\right) \Gamma\left(1+\frac{\ell}{2} - i\frac{\omega}{2}q + 2ia_L\right)}{\Gamma\left(-\frac{\ell}{2} - i\frac{\omega}{2}q + 2ia_L\right) \Gamma\left(-\frac{\ell}{2} - i\frac{\omega}{2}p + 2ia_R\right)}.
		\end{equation}
		
		The imaginary and real parts of these coefficients are, respectively,
		\begin{align}
			&\text{Im}(k_{\ell}) = -\mathcal{K}_{\ell} \sin(\pi\ell)\sinh\left(\pi(y_1+y_2)\right) \hspace{15pt} \label{dissipativeGen} \\ & \hspace{-40pt}\text{and}  \nonumber \\
			&\text{Re}(k_{\ell}) =\mathcal{K}_{\ell} \left[ \cosh\left(\pi(y_1-y_2)\right) - \cos(\pi\ell)\cosh\left(\pi(y_1+y_2)\right) \right] \label{ConservativeGen},
		\end{align}
		
		\noindent where $y_1$ and $y_2$ are defined as before and we have introduced the real constant
		\begin{equation}
			\mathcal{K}_{\ell} \equiv \frac{(r^2_{+} - r^2_{-})^{1+\ell} }{2\pi^2}\frac{\Gamma(-1-\ell)}{\Gamma(1+\ell)}\left|\Gamma\left(1+\frac{\ell}{2} + iy_1\right)\right|^2 \left|\Gamma\left(1+\frac{\ell}{2} + iy_2\right)\right|^2.
		\end{equation}
		Since the Gamma functions have no zeroes, $K_{\ell}$ is nonzero for all choices of parameters.
		
\begin{figure}[h]
			\centering
			\includegraphics[width=8.2cm]{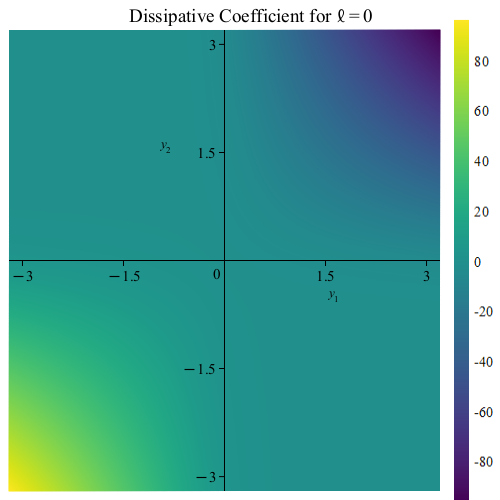}
			\hspace{0.25cm}\includegraphics[width=8.75cm, height=8.2cm]{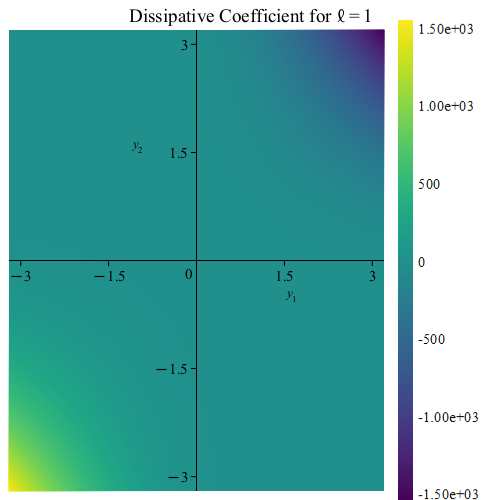}
			\caption{Contour plots representing the dissipative part of response coefficients for the $\ell=0$ (left) and $\ell=1$ (right) cases, as functions of $y_1$ and $y_2$.}
			\label{Dissipative}
		\end{figure}

		We may now send $\ell$ to its physical integer values. The expressions above are undefined for integer $\ell$, but may be analytically continued by letting $\ell=n+\epsilon$, with $n\in\mathbb{N}$, and evaluating the limit as $\ell\to\infty$. Naively, one may expect that $\text{Im}(k_{\ell})\to 0$ in this limit, as the sine function goes to zero. However, $\Gamma(-1-\ell)$ also develops a simple pole at $\ell\in \mathbb{N}$, so we must evaluate the limit carefully. In fact, calculation of the residue of the Gamma function shows that $\Gamma(-1-\ell)\approx \frac{(-1)^{\ell + 1}}{\epsilon (\ell+1)!}$, while $\sin(\pi\ell)\approx (-1)^{\ell}\pi\epsilon$ at the same limit. Thus, the physical dissipative coefficients of the STU black hole are given by
		\begin{equation}\label{dissipativecoeff}
			\text{Im}(k_{\ell}) = - \frac{(r^2_{+} - r^2_{-})^{1+\ell}}{2\pi \ell!(\ell+1)!} \sinh\left(\pi(y_1+y_2)\right) \left|\Gamma\left(1+\frac{\ell}{2}+iy_1\right)\right|^2 \left|\Gamma\left(1+\frac{\ell}{2}+iy_2\right)\right|^2,
		\end{equation}
		which are finite, and in general nonzero, parameters, indicating a well-defined dissipative response. In Fig.~\ref{Dissipative}, we display the functional dependence of these dissipative coefficients on $y_{i}$, for two fixed values of $\ell$.

		The real part of the tidal coefficients, which is identified with the Love numbers, behave as  
		\begin{equation}\label{LoveNumbers}
			\text{Re}(k_{\{N+\epsilon\}})\xrightarrow[\varepsilon \to 0]{} \frac{R_{N}}{\epsilon}
		\end{equation}
		when $ \epsilon \to 0$, therefore displaying poles of order one as $\ell$ approaches its physical values. The finite residue of this pole, $R_{\ell}$, $\ell\in\mathbb{N}$, is calculated to be:
		\begin{equation}
			\begin{split}
				R_{\ell} = &\frac{(-1)^{\ell} (r^2_{+} - r^2_{-})^{1+\ell}}{2\pi^2 \ell! (\ell+1)!} \left[ \cosh(\pi(y_1-y_2)) - (-1)^{\ell}\cosh(\pi(y_1+y_2)) \right]\\ & \ \ \times\left|\Gamma\left(1+\frac{\ell}{2}+iy_1\right)\right|^2 \left|\Gamma\left(1+\frac{\ell}{2}+iy_2\right)\right|^2 .
			\end{split}
		\end{equation}
		
		Thus, we see that Love numbers are divergent unless $R_{\ell}$ has a zero of at least order one, which is not true for generic values of frequency and model parameters. Such divergences are expected, as they are associated with the well-known presence of classical Renormalization Group (RG) running in the effective field theory interpretation. Despite this pole, the total wavefunction calculated above is regular at infinity because of a compensating divergence from the source part of the system. These results are consistent with the known literature for neutral black holes and in fact become equivalent to that found in Ref.~\cite{Charalam_2023} when all $\delta_i\to 0$. 
		
		Note that in the static case, which is achieved by simply setting $y_1=2a_R$ and $y_2=2a_L$ in expressions~\eqref{dissipativecoeff} and~\eqref{LoveNumbers}, the vanishing of Love numbers again  requires even $\ell$ as well as equality in the magnitude of the hyperbolic cosine arguments in~\eqref{LoveNumbers}, i.e,
		$a_R-a_L=\pm\left(a_R+a_L\right)$. Thus, we again find that the vanishing of static Love numbers requires the fulfillment of conditions~\eqref{StaticLNvanishingcond}. however, note that the static dissipative coefficients can only vanish if $a_R+a_L=0$. These conditions cannot be consistent with~\eqref{StaticLNvanishingcond} unless \emph{both} $a_R$ and $a_L$ are zero, so that static dissipative effects generally persist even when the conservative part of the tidal response disappears. As we mentioned before, static solutions satisfying~\eqref{StaticLNvanishingcond} have precisely the same form as their $4D$ rotating black hole analogues with $\frac{\ell}{2}$ replacing $\ell$ in the $5D$ theory. Thus, the emergence of purely dissipative tidal responses in these cases is consistent with the known results for Kerr and other rotating black holes ~\cite{Chia, LeTiec, LoveSymmetry}, which have also been verified to hold for the four-dimensional STU black holes~\cite{CveticPRD105}.
		
		Clearly, the response coefficients $k_{\ell}$ found through analytic continuation are not equivalent to the $\lambda_{\ell}$ given in Eq.~\eqref{Love_n}, as the latter are real-valued, bounded parameters for all finite values of $y_{1}$ and $y_2$. This difference is manifest even in the static case, where  the analytic continuation method gives the response coefficients
		\begin{equation}\label{klstatic}
			\left.k_{\ell}\right|_{\omega=0} = \vk^{1+\ell} \frac{\Gamma(-1-\ell)}{\Gamma(1+\ell)} \frac{\Gamma\left(1+\frac{\ell}{2}  + 2ia_R\right) \Gamma\left(1+\frac{\ell}{2}  + 2ia_L\right)}{\Gamma\left(-\frac{\ell}{2} + 2ia_L\right) \Gamma\left(-\frac{\ell}{2} + 2ia_R\right)},
		\end{equation}
		which are quite different from the static response coefficients~\eqref{Love_n_st} obtained in the previous section. In particular, we see that the dissipative coefficients, obtained by setting $y_{1/2}=2a_{R/L}$ in Eq.~\eqref{dissipativecoeff} are generally nonzero even in this limit. These results were in fact expected, and also found for the previously studied neutral black holes, as they are a consequence of source-response ambiguity. Our mathematical treatment introduces a source at infinity and, through the required boundary conditions, ensures that it produces an $r^{\ell}$ tidal potential asymptotically, which causes a tidal response in the perturbative solutions. In the Newtonian scenario, this leads to a tidal potential described, at large distances, by a linear superposition of these two contributions. Once relativistic corrections are introduced, such a clear separation ceases to exist, and it can be verified that the source and response asymptotic expansions overlap due to emergence of terms with exactly the same radial dependence in both series. Mathematically, this ambiguity is manifest in the fact that the radial equation falls within a special case of the hypergeometric equation, which is why recourse to the series expansion~\eqref{HG_abn} is necessary and why logarithmic contributions must arise. 
		
		The analytical continuity approach resolves the source-response degeneracy by exploiting the fact that the Frobenius' series overlap does not arise for non-integer $\ell$. Thus, we may associate the coefficients found in this way to the black hole's intrinsic deformability, stripped of any confounding contributions from the source field. The fundamental, gauge-invariant Love numbers are thus identified with the conservative part of $k_{\ell}$. This interpretation is also strongly supported by the EFT approach, where it can be shown that these parameters can be interpreted in terms of Wilson coefficients~\cite{EFT1,EFT2, EFT3,EFT4}. For these reasons, and aiming to avoid confusion, we shall use the term ``Love numbers" exclusively in reference to the real part of $k_{\ell}$, although the alternative approach, namely calling the real coefficients $\lambda_{\ell}$ Love numbers, is also valid and found throughout the literature.
		
		Note that the above discussion does not lessen the importance of the coefficients~\eqref{Love_n}. In fact, $\lambda_{\ell}$ is a composite quantity which combines the contributions of $k_{\ell}$ and the ambiguity terms. Unlike $k_{\ell}$, the $\lambda_{\ell}$ are finite quantities, because the response divergence is exactly compensated by an equivalent source contribution, leaving out a finite part expressed by the composite coefficient $\lambda_{\ell}$. The logarithmic term in~\eqref{Love_n} introduces a scale dependence that signifies classical Renormalization Group flow. In the EFT point of view, this can be quantified trough the introduction of a matching scale $\mu$~\cite{BHstereotyping}, so that $k_{\ell}\propto \ln(\vk/\mu^2)$ hence resulting in a beta function $\mu\frac{dk_{\ell}}{d\mu}\propto \lambda_{\ell}/\ln\left[\vk/(r^2-r_{-}^2)\right]$. Thus, the combined information provided by $k_{\ell}$ and $\lambda_{\ell}$ gives a more complete description of the manner in which the divergences are canceled due to source-response ambiguity, and of the manifestation of RG flow. These quantities hence play complementary roles in the theory, and together paint a clearer picture of the black hole tidal problem. An interesting observation is that, although the $k_{\ell}$ are in principle the quantities whose vanishing is connected to the highest-weight representation derived from Love symmetry, we find that the $\lambda_{\ell}$ also vanish in the same circumstances, so the conditions derived imply the vanishing of the conservative part of \emph{both} tidal coefficients.

		Another important observation is that, unlike what we found in the static case, the dynamical tidal responses are \emph{not} described solely in terms of seed parameters from the Myers-Perry solution, meaning that the U(1) charges leave a distinctive signature in these parameters, as seen from~\eqref{Def:p_and_q}. We therefore see from the results of this section that Love numbers are indeed changed continuously when the boost parameters $\delta_i$ (and thus the physical charges) are varied. Here we observe a novel feature of the STU solution in relation to the more well-known neutral black holes, as the boost-dependent variables $p$ and $q$ are seen to be as important the frequency, which only appears in the response coefficients through the products $p\omega$ and $q\omega$, in the dynamical tidal description. The charge dependence vanishes completely as $\omega\to 0$, thus agreeing with our previous observation that static tidal responses can be completely described in terms of the seed parameters of the Myers-Perry solution.

		\subsection{Special cases: Vanishing dynamical Love numbers/ dissipative coefficients}
		
		Before ending this section, let us consider a few special cases. First, let us see if there are specific choices of the model parameters such that the Love numbers are finite. The divergence in the conservative response comes from a first-order pole that arises in $\Gamma(-1-\ell)$ as $\ell$ approaches a natural number, so it can be regularized by a zero in the numerator. Since $K_{\ell}$ in nonzero, the pole can only be canceled if the following equality holds:
		\begin{equation}
			\cosh(\pi(y_1-y_2)) = (-1)^{\ell}\cosh(\pi(y_1+y_2)).
		\end{equation}
		
		The fact that $\cosh(x)$ is strictly positive for real arguments precludes solutions of this equation for odd $\ell$. If, however, $\ell$ is an even number, the parity of the hyperbolic sine implies equality is attained if and only if one of two conditions hold, namely,
		\begin{align}\label{DynamicalLNVanishingGen}
			(y_1-y_2) &= (y_1+y_2) &&\implies& a_R = \frac{\omega}{4}p \\
			\text{or} \hspace{60pt}& \nonumber\\ \quad (y_1-y_2) &= -(y_1+y_2) &&\implies& a_L = \frac{\omega}{4}q.
		\end{align}
		
		If either of these equalities is satisfied, the numerator of~\eqref{LoveNumbers} has a zero of order two, so that Love numbers are not only finite, but in fact vanish. In the static case, these are the vanishing conditions~\eqref{StaticLNvanishingcond} discussed before, which are extended to the dynamical case by replacing the vanishing of $a_{R}$ $(a_L)$ in those conditions by that of $y_{1}$ $(y_2)$. These conditions would generally only be attained by perturbations of a given frequency, whose value depends on the specific values of the parameters, but there are a few interesting special cases. Among these, we highlight two possibilities:
		\begin{align}\label{DynamicalVanishingCond}
			a_R=p=0   &&  \text{ and } &&   a_L=q=0,
		\end{align}
		The above conditions imply Love number vanishing for \emph{all} frequencies in the even $\ell$ cases. In this sense, they generalize the $a_{L/R}=0$, $\ell=2n$ conditions of the static equation to the dynamical scenario within the near-zone regime. Note that definitions~\eqref{Def:p} and~\eqref{Def:q} imply that, for nonzero angular momenta, these conditions can only be satisfied by charged black holes. Indeed, by  setting $\delta_{i}=0$ in the definitions of $p$ and $q$, we see that these parameters vanish only if $\mu\to 0$. But in order for $\Delta$, and thus the physical horizons, to be real-valued, the metric must be such that $\mu \geq (|l_1|+|l_2|)^2$, so that the imposition of conditions~\eqref{DynamicalLNVanishingGen} to a neutral solution leads to Schwarzschild-Tangherlini black hole with vanishing mass. For charged black holes, there is however no such restriction, as the boost parameters must also be taken into account, and they in fact make it possible for $p$ or $q$ to vanish without forcing $M$ or the angular momenta to be zero. Indeed, conditions~\eqref{DynamicalVanishingCond} are attained for rotating, charged black holes if 
		\begin{equation}\label{tangentEq}
			\prod_{i=1}^{3}\cosh\delta_i=\pm\prod_{i=1}^{3}\sinh\delta_i \implies \prod_{i=1}^{\bar{j}\leq 3}\tanh(\delta_i) =\pm 1,
		\end{equation}
		where $\bar{j}$ represents the number of nonzero $\delta_i$ (which is simply three if the black hole has all three charges), and the upper and lower signs lead, respectively, to the vanishing of $p$ and $q$. Since $|\tanh(\delta_i)|<1$, the above equations cannot be solved by any finite choice of the parameters. However, equality is clearly attained in the limit $\delta_i \to \pm \infty$. If all nonzero $\delta_i$ have the same sign, then $p\to 0$, while different signs imply $q\to 0$. If the limits $\delta_i \to \pm \infty$ are taken literally and the other parameters remain finite and arbitrary, then we must conclude that the physical $U(1)$ charges also go to infinity. However, dynamical Love numbers are defined only within the limits of the near-zone regime, so in fact we need only consider $\delta_i$ large enough to ensure that either $\omega p$ or $\omega q$ is negligible to the desired level of approximation. That said, we shall shortly demonstrate that it is also possible to satisfy these conditions exactly through saturation of the Bogomol'nyi bound.
		
		Let us exemplify these results with arguably the simplest charged possibility, namely that of a solution with only one electric charge. If $|\delta_1| >>1 $ while $\delta_{2}=\delta_3=0$, $p$ and $q$ behave as
		\begin{equation}\label{p_q_KN}
			p,q\sim \mu\,\dfrac{\exp(|\delta|)\left(1\pm \frac{Q}{|Q|}\right)}{\sqrt{\mu - \left(l_{1} \pm l_{2}\right)^{2}}}  \ \ \ \ \text{(5D one-charged solution)}
		\end{equation}
		so that $q\sim 0$ if $Q$ has a positive sign, while $p$ is the one that is small if $Q<0$. Under these conditions, the hypergeometric parameter $\ma$ becomes a positive integer for even $\ell$, hence falling into the degenerate case and ensuring the polynomial form typical of vanishing Love numbers, while frequency dependence of the polynomial coefficients is encoded within the parameters $\mb$ and $\mc$ which both contain terms proportional to $i\omega Q^{1/2}$. When $\omega\to 0$, we recover the polynomial coefficients of the static case.

		In general, the vanishing  condition depends on the sign combination of all nonzero charges, as the boosts can go to plus or minus infinity. The condition thus becomes
		\begin{equation}\label{gen}
			p,q\sim \mu\frac{\exp(\sum_{i=1}^{3}|\delta_i|)}{\sqrt{\mu - \left(l_{1} \pm l_{2}\right)^{2}}}\left[1\pm \text{sign}\left(\prod_{i=1}^{\bar{j}\leq 3}\delta_i\right)\right].
		\end{equation}
		
		A particularly important scenario in which conditions~\eqref{DynamicalVanishingCond} are found to be (nearly) met is that of (near-)Bogomol'nyi saturation. BPS solutions possess the minimal mass compatible with the given charges and angular momenta, and are specially important in supersymmetric settings due to partial conservation of supersymmetry. As explained above, saturation of the Bogomol'nyi bound in STU black holes is achieved by simultaneously sending $\mu$ to $0$ and letting the boosts $\delta_i$ grow to infinity, while keeping the charges and angular momenta fixed as limiting expressions of~\eqref{charges} and~\eqref{eq:l12def}. Because $\mu\to 0$ creates a double root in the horizon equation, the coordinate transformation~\eqref{eq:xdef} is not well-defined in the BPS limit, but still applies to black holes that are arbitrarily close to saturating the Bogomol'nyi bound. The defining requirements for physically acceptable near-BPS solutions are~\cite{cy96a}:
		\begin{align}\label{BPS}
			|l_1|\sim |l_2|\sim \mu^{1/2} && |Q_i|\sim \mu\exp(2\delta_i)\sim M ,
		\end{align}
		and we may for the moment assume $\mu> |l_1|^2 + |l_2|^2$ to prevent extremality or  naked singularities. When~\eqref{BPS} are satisfied, the parameters $p$ and $q$ behave as 
		\begin{equation}\label{genBPS}
			p,q\sim \mu^{1/2}\exp\left(\sum_{i=1}^{3}|\delta_i|\right)\left[1\pm \text{sign}\left(\prod_{i=1}^{3}\delta_i\right)\right].
		\end{equation}
		
		We note that, although relations~\eqref{gen} are still formally valid as approximations if some of the boosts vanish, the corresponding BPS black holes only exist if all charges are nonzero, which is why we have set $\bar{j}=3$ in~\eqref{genBPS}. This can be seen from the horizon surface area, which in the BPS limit can be shown to satisfy $A_{BPS}\propto\sqrt{Q_1Q_2Q_3}$~\cite{cy96a}, and thus vanishes unless all charges are nonzero. If $\mu$ is small but nonzero, we can expand the thermodynamic parameters in terms of $\mu$ to find, in particular, that this surface area gains a linear correction $\mathcal{O}(\mu)$, so that~\eqref{genBPS} works as an approximation in this sense. In the one-charged black hole example used above, setting $|l_{1/2}|^2\sim\mu << 1$ would allow us to formaly write $p,q \sim \mu^{1/2}e^{|\delta|}\left(1\pm \text{sign}(Q)\right)$, leading to a non-BPS black hole with $q\approx 0$, $p\approx 4\sqrt{2}Q$ (assuming $Q>0$).

		From Eq.~\eqref{genBPS} we see that as the parameters of the model approach the Bogomol'nyi limit either $p$ or $q$ must vanish, so that when $a_{R/L}=0$, conditions~\eqref{DynamicalVanishingCond} are always (nearly) attained as a consequence of (near) BPS saturation, with no further hypotheses needed. We should however note that the surviving parameter in~\eqref{gen} may grow exponentially with $\delta$. In fact, if $q$ goes to zero, we see that 
		$p\propto Q_1\exp(|\delta_2|+|\delta_3|)$ in the near-BPS case, growing exponentially with $\delta_i$. An analogous behavior is encountered for $q$ if the sign of the boost product in~\eqref{gen} is flipped. These parameters would thus diverge if \emph{exact} BPS saturation is assumed, as we must send $\delta_i$ to infinity in that limit. This is most likely just a result of the fact that the coordinate transformation~\eqref{eq:xdef} becomes singular when $\Delta\to 0$. The exponential growth in $q$ is properly compensated by this coordinate singularity, and does not seem to affect the observable thermodynamic quantities, which have a well-defined, finite value in the BPS case.
		
		Let us consider this last issue more carefully. For definiteness, assume $\delta_i>0$ for all $i$ and let these boosts go to infinity while $\mu$, $|l_1|$ and $|l_2|$ go to zero in a manner consistent with~\eqref{BPS}. Along with~\eqref{genBPS}, these conditions imply $q\to 0$ and
		\begin{equation}\label{pBPS}
			p\sim 2\mu^{1/2}e^{\delta_1+\delta_2+\delta+3}\sim\frac{2}{\mu}(8Q_1Q_2Q_3)^{1/2},
		\end{equation}
		which thus diverges as $1/\mu$ as the BPS limit is approached. Now, as $\mu\to 0$ we see from~\eqref{Defrprm} and~\eqref{Delta} that $r_{\pm}$ become equal, so that the near-zone equation becomes a confluent hypergeometric differential equation. The response coefficients thus obtained can still be verified to agree with the real part of~\eqref{kl} in the confluent limit. As $p\to\infty$, a singularity emerges from the $p$-dependent terms in these expressions, while the other Gamma functions, being independent of this parameter, remain unchanged. We must thus consider the asymptotic ratio
		\begin{equation}
			\frac{\Gamma\left(1+\frac{\ell}{2}- i\frac{\omega}{2}p + 2ia_R\right)}{\Gamma\left(-\frac{\ell}{2}- i\frac{\omega}{2}p + 2ia_R\right)} \sim \left(-i \frac{\omega}{2} p\right)^{1+\ell}¨
		\end{equation}
		which, up to a finite, but otherwise irrelevant phase factor, behaves as $|\omega |p^{1+\ell}/2$ as conditions~\eqref{BPS} are approached. By~\eqref{pBPS}, this amounts to a pole of order $\ell + 1$ in $\mu$. However, substitution of~\eqref{BPS} into~\eqref{Defrprm} and~\eqref{Delta} shows that $\vk\sim \mu$ as the parameters approach the BPS conditions. Thus, the zero from $\vk^{\ell +1}$ exactly compensates the pole generated by $p$, so that $k_{\ell}$ remains well-defined in this limit. From the above considerations, it seems reasonable to extend the near-BPS results discussed above to the limit of Bogomol'nyi saturation, in which conditions~\eqref{DynamicalVanishingCond} are \textit{exactly} satisfied. Conversely, this is also the only scenario in which this exact fulfillment of~\eqref{DynamicalVanishingCond} can be achieved while keeping the mass and charges finite, so one is naturally led to~\eqref{BPS} upon assuming these conditions, without use of supersymmetric arguments. 
		
		The results discussed in this subsection also have a thermodynamic meaning. In fact, by comparing the definitions of $p$ and $q$ to ~\eqref{eq:betar} and \eqref{eq:betal}, we see that these parameters are respectively proportional to the right and left inverse temperatures, so that the vanishing of dynamical Love numbers follows from the vanishing of one of either $\beta_{R}$ or $\beta_L$. By the same token, we see from~\eqref{eq:l12def},~\eqref{eq:SL} and~\eqref{eq:SR} that these limits also imply the vanishing of either left or right angular momenta and entropies. Note that, if we limit ourselves to approximate results, these properties \emph{do not} imply saturation or near-saturation of the BPS bound; the weaker assumption $|\delta_i|>>1$ with finite $\mu$ suffices to ensure the vanishing of one of these quantities to the desired level of approximation. Although not necessary, the BPS limit is certainly sufficient to ensure the vanishing of these thermodynamic quantities, and the hierarchy between left and right parameters for BPS solutions of the STU family has been observed before~\cite{cy96a}. Moreover, this is also the only physical limit in which the vanishing of these quantities can be exact, as in other scenarios the $\delta_i$ must remain bounded to prevent infinite charges.
		
		The regime in which conditions~\eqref{DynamicalVanishingCond} are satisfied is also important from a string-theoretical point of view. It has long been conjectured that the entropy of black hole solutions from string theory can be derived from the counting of solitonic states associated with  D-brane configurations (see Refs.~\cite{Strominger, DbranecountingII, Dbranecounting, DbranecountingIII} for more information on the topic). This relationship is of great theoretical significance, and while it has been mostly studied in the Bogomol'nyi limit or near it, investigations of this issue also exist outside of BPS saturation. In Refs.~\cite{cl97a, Larsen}, the authors proposed an explanation of the STU black hole entropy as the sum of contributions from left and right moving modes of the string model. The proposed levels of the effective string are~\cite{cl97a}
		\begin{align}
			N_L &= \frac{1}{4}\,\mu^3
			\bigl(\Pi_c + \Pi_s\bigr)^2
			- J_L^2, \\[6pt]
			N_R &= \frac{1}{4}\,\mu^3
			\bigl(\Pi_c - \Pi_s\bigr)^2
			- J_R^2, 
		\end{align}
		so that the entropy in the horizon can be written as the sum $S = S_L + S_R = 2\pi\bigl(\sqrt{N_L} + \sqrt{N_R}\bigr).$ Since, as remarked above, conditions~\eqref{DynamicalVanishingCond} imply the vanishing of either $J_R$ or $J_L$, the vanishing of these levels follows from $\Pi_c=\pm \Pi_s$ (if $\mu$ is finite), which implies $p=0$ or $q=0$, whether or not BPS saturation is assumed. We see that the vanishing of dynamic Love numbers for any frequency corresponds to the important limiting case of a black hole whose thermodynamic properties are entirely determined by either left or right moving modes.

		\cmmnt{In particular, the fact that angular momenta $J_{\phi}$ and $J_{\psi}$ are kept fixed in the BPS limit implies that there exist $L_1$ and $L_2$ such that 
			\begin{align}
				J_{\phi}=\sqrt{2\prod_{i=1}^{3}Q_i}\left(L_1\pm L_2\right) =\pm J_{\psi},
			\end{align}
			so that either $J_R$ or $J_L$ must vanish in the BPS limit.  }

		Now, let us investigate the conditions that lead to vanishing dissipative coefficients, meaning that the tidal response of the black hole is strictly conservative. First, note that, since $\mathcal{K}_{\ell}\neq 0$, the dissipative coefficients can only vanish if the hyperbolic sine vanishes, meaning that we must have $y_1+y_2 =0$. In terms of the parameters of the model, this equality in translated into the condition
		\begin{equation}
			a_L+a_R=\frac{\omega}{2\kappa_{+}},
		\end{equation}
		where use was made of the fact that $p+q=2/\kappa_{+}$. Although this constraint is in general charge-dependent because $\kappa_{+}$ is a function of the $\delta_i$, the condition has the same form as in the Myers-Perry case, which is identified with the $\delta_{i}\to 0$ limit of this expression. The generalization to charged black holes manifests itself in the fact the frequency needed to satisfy this equation changes by a factor of $\kappa_{+}/\tilde{\k}_{+}$, where $\tilde{\k}_{+}$ is the surface acceleration from the Myers-Perry solution. By recalling the definitions of $a_L$ and $a_R$, we could alternatively write this expression in the form
		\begin{equation}\label{DissVanishingCond}
			\omega=2\left(m_R\Omega^R_{+} + m_L\Omega^L_{-}\right),
		\end{equation}
		
		If $\omega=0$, the above equation is reduced to $m_R\Omega^R_{+}=-m_L\Omega^L_{+}$, or $a_R=-a_L$. This is achieved by Schwarzschild-Tangherlini black holes and charged black holes generated from it, as well as rotating black holes with nonzero but opposite $a_L$ and $a_R$. In general, Eq.~\eqref{DissVanishingCond} indicates that, for each combination of the model parameters and quantum numbers, there exists a unique frequency for which the tidal responses are completely conservative. 
		
		Interestingly, the above condition is precisely equivalent to removal of the outer horizon singularity from the near-zone equation~\eqref{eq.nzeq}, since it makes the term proportional to $(\bx-1/2)^{-1}$ vanish. We thus see that dissipation of the tidal response can be physically understood as the result of the interaction between perturbative modes and the event horizon of the black hole, which makes physical sense since absorption at the horizon can account for dissipative effects. 
		
		\subsection{Complex-valued frequencies}
		As remarked above, the calculation of the imaginary and real parts in Eqs.~\eqref{dissipativeGen} and~\eqref{LoveNumbers} assumes $\omega$ real, although~\eqref{kl} remains valid in the complex plane. As argued, we are most interested in the real-valued frequencies from which can we can read-off the influence of a stable, persistent tidal field, rather than transient effects that vanish exponentially fast. However, complex frequencies are important in some contexts such as quasinormal mode investigation, so it is worthwhile to present, in the interest of completeness, the generalization of~\eqref{dissipativecoeff} and~\eqref{LoveNumbers} to the scenario of complex-valued frequencies. To this end, let $\omega = \omega_{\mathrm{Re}} + i \omega_{\mathrm{Im}}$ and let us introduce the parameters
		\begin{align}
			y_1 &= \left(2a_R - \frac{\omega_{Re}p}{2}\right) - i\left(\frac{\omega_{Im}p}{2}\right) \equiv y_1^{Re} + iy_1^{Im} \\[1.5ex]
			y_2 &= \left(2a_L - \frac{\omega_{Re}q}{2}\right) - i\left(\frac{\omega_{Im}q}{2}\right) \equiv y_2^{Re} + iy_2^{Im}
		\end{align}
		
		The expression we found for $k_{\ell}$ is still valid when written in terms of $y_1$ $y_2$, but now the fact that these are complex variables causes extra terms to appear when the real and imaginary parts are extracted. Using addition formulae to expand the $\omega$-dependent hyperbolic functions in terms of $ \omega_{\mathrm{Re}}$ and $\omega_{\mathrm{Im}}$, we find

		\begin{equation}\label{LoveNumbersComplex}
			\begin{split}
				\operatorname{Re}(k_\ell)= \mathcal{K}_{\ell} &\Bigg[ 
				\cosh\left( \pi (y_1^{\mathrm{Re}} - y_2^{\mathrm{Re}}) \right) \cos\left( \pi (y_1^{\mathrm{Im}} - y_2^{\mathrm{Im}}) \right) \\
				&\quad - \cos(\pi \ell) \cosh\left( \pi (y_1^{\mathrm{Re}} + y_2^{\mathrm{Re}}) \right) \cos\left( \pi (y_1^{\mathrm{Im}} + y_2^{\mathrm{Im}}) \right) \\
				&\quad + \sin(\pi \ell) \cosh\left( \pi (y_1^{\mathrm{Re}} + y_2^{\mathrm{Re}}) \right) \sin\left( \pi (y_1^{\mathrm{Im}} + y_2^{\mathrm{Im}}) \right) \Bigg]
			\end{split}
		\end{equation}
		and 
		\begin{equation}
			\begin{split}
				\operatorname{Im}(k_\ell) = \mathcal{K}_{\ell} &\Bigg[ 
				\sinh\left( \pi (y_1^{\mathrm{Re}} - y_2^{\mathrm{Re}}) \right) \sin\left( \pi (y_1^{\mathrm{Im}} - y_2^{\mathrm{Im}}) \right) \\
				&\quad - \cos(\pi \ell) \sinh\left( \pi (y_1^{\mathrm{Re}} + y_2^{\mathrm{Re}}) \right) \sin\left( \pi (y_1^{\mathrm{Im}} + y_2^{\mathrm{Im}}) \right) \\
				&\quad - \sin(\pi \ell) \sinh\left( \pi (y_1^{\mathrm{Re}} + y_2^{\mathrm{Re}}) \right) \cos\left( \pi (y_1^{\mathrm{Im}} + y_2^{\mathrm{Im}}) \right) \Bigg]
			\end{split}		
		\end{equation}
		
		The above expressions are reduced to Eqs.~\eqref{dissipativeGen} and~\eqref{LoveNumbers} when $\omega_{\rm{Im}}=0$, and are natural generalizations of these expressions to the case of complex $\omega$. The coefficients in~\eqref{Love_even} are not proper analogues of the Newtonian Love numbers, being instead related to ringdown responses caused on the black hole by time-decaying perturbations. These may, however, be measurable by gravitational wave detectors, and thus such responses may offer insight into the black hole's internal structure. 
		
		We can also investigate the conditions that lead to absence of RG flow and lack of dissipative part, just as we did in the case of real $\omega$. For the real part of the conservative response to be non-divergent, the following condition must be met:
		\begin{equation}\label{CondNDReal}
			\cosh\left( \pi (y_1^{\mathrm{Re}} - y_2^{\mathrm{Re}}) \right) \cos\left( \pi (y_1^{\mathrm{Im}} - y_2^{\mathrm{Im}}) \right) \\ 
			= (-1)^\ell \cosh\left( \pi (y_1^{\mathrm{Re}} + y_2^{\mathrm{Re}}) \right) \cos\left( \pi (y_1^{\mathrm{Im}} + y_2^{\mathrm{Im}}) \right).
		\end{equation}
		
		On the other hand, the imaginary part vanishes if and only if
		\begin{equation}\label{CondVanishingIm}
			\sinh\left( \pi (y_1^{\mathrm{Re}} - y_2^{\mathrm{Re}}) \right) \sin\left( \pi (y_1^{\mathrm{Im}} - y_2^{\mathrm{Im}}) \right) \\
			= (-1)^\ell \sinh\left( \pi (y_1^{\mathrm{Re}} + y_2^{\mathrm{Re}}) \right) \sin\left( \pi (y_1^{\mathrm{Im}} + y_2^{\mathrm{Im}}) \right)
		\end{equation}
		
		These conditions are far less strict then their real $\omega$ analogues. In fact, both the even and odd versions of these equations have an infinitely large manifold of solutions. As an example, let us consider the case  $\ell=2n+1$, for which the real $\omega$ Love numbers were always divergent due to the impossibility of zeroes in the numerator of~\eqref{LoveNumbers}. For complex-valued frequencies, solutions exist if, for example, both sides are zero. This happens if $\cos\left( \pi (y_1^{\mathrm{Im}} - y_2^{\mathrm{Im}}) \right) = 0$ and $\cos\left( \pi (y_1^{\mathrm{Im}} + y_2^{\mathrm{Im}}) \right) = 0$, which implies 
		\begin{align}
			y_1^{\mathrm{Im}}=\frac{j+k}{2}, && \text{and} \hspace{90pt} y_2^{\mathrm{Im}}=\frac{j-k}{2},
		\end{align}
		where $j$ and $k$ are odd integers. For even $\ell$, a larger class of solutions (which is still a small subset of the general case) is achieved by setting $y_1^{\mathrm{Im}}=	y_2^{\mathrm{Im}}$, in which case the simple condition $	\cosh\left( \pi (y_1^{\mathrm{Re}} - y_2^{\mathrm{Re}}) \right)=	\cosh\left( \pi (y_1^{\mathrm{Re}} + y_2^{\mathrm{Re}}) \right)\cos(2\pi y_1^{\mathrm{Im}}) $, which can be satisfied if the right-hand side is greater than or equal to 1, gives a continuous family of solutions parameterized by $y_1^{\mathrm{Im}}$. The case $y_1^{\mathrm{Im}}=0$ gives precisely the conditions that lead to Love number vanishing as discussed in the previous subsection. 
		
		Some interesting, if fine-tuned, scenarios also exist regarding the vanishing of the dissipative part. One way that a purely real response can be achieved is if the real part of one of the parameters, say $y_2^{\mathrm{Re}}$, vanishes along with a constraint on the imaginary parameters. If $\ell$ is even, the constraint is  
		$\sin\left( \pi (y_1^{\mathrm{Im}} - y_2^{\mathrm{Im}}) \right) = \sin\left( \pi (y_1^{\mathrm{Im}} + y_2^{\mathrm{Im}}) \right)$. In the odd case, the simpler condition $y_1^{\mathrm{Im}}=0$ is sufficient. Thus a ``dissipationless" state is encountered if one of the $y_i$ is purely real while the other is purely imaginary, which is equivalent to the vanishing of either $p$ or $q$ along with a constraint involving $\omega_{Re}$. If, for example, $\omega_{Re}=0$ (that is, a purely dissipative frequency), then the condition becomes $p=0$ ($q=0$) along with $a_L=0$ ($a_R=0$) with odd $\ell$, which coincidentally makes up for a perfect complement of conditions~\eqref{DynamicalVanishingCond} discussed in the previous subsection, and which were valid for even $\ell$ and real frequencies. 
		
		The special examples discussed in this subsection appear somewhat fine-tuned, and their physical relevance is thus unclear at the moment. Nevertheless, they offer some interesting insight into the mathematical structure of the theory, and are used to illustrate the many ways in which~\eqref{CondNDReal}and \eqref{CondVanishingIm} may be satisfied. Moreover, the response coefficients discussed in this subsection may be useful in discussions related to quasinormal modes, where they may provide high precision tests to gravitational theories.

		\section{Love symmetry of the near-zone scalar equation}\label{sec:LoveSymmetry}
		
		The vanishing of static Love numbers (generally in 4D an for selected cases in 5D) has been explained in terms of selection rules that follow from a hidden ``Love" symmetry of the near-zone equation~\cite{LoveSymmetry, Charalam_2023}. This symmetry is found in several black hole solutions, and has been linked to the near horizon isometries of extremal black holes, a remnant of which is still manifest within the algebra of $SL(2,\mathbb{R})$  that characterizes Love symmetry. In this framework, it is found that solutions with vanishing Love numbers correspond states belonging to a highest-weight representation implied by the Love symmetry generators. In this section, we shall derive these generators by imposing both the algebra and identification between the Casimir and near-zone equations, and investigate the consequences of this symmetry.
		
		In the near-zone limit we are able to introduce three vector fields $\hat{L}^{(\sigma)}_{m}$ satisfying the $sl(2,{\mathbb{R}})$ algebra commutation relations, namely:
		\be 
		\left[\hat{L}_m,\hat{L}_{n}\right]=(m-n)\hat{L}_{m+n}, \quad m,n=0,\pm.\label{SL2cr}
		\ee
		The Casimir ${\cal {\hat C}}_{2}$ of the algebra generated by the $\hat{L}_m$ corresponds to the differential operator $\hat{\cal D}$ that gives rise to the near-zone wave equation (\ref{eq.nzeq}). We have:
		\be
		{\cal {\hat C}}_{2}=\hat{L}^2_{0}-{1\over 2}\left(\hat{L}_{+1}\hat{L}_{-1}+\hat{L}_{-1}\hat{L}_{+1}\right),\label{Casimir}
		\ee
		for both signs $\sigma=\pm 1$.  The operator $\hat{\cal D}$ is of the form:
		\be
		\hat{\cal D}=-{\p\over{\p\bx}}\left({\bx}^2-{1\over 4}\right){\partial\over\partial \bx}-{{\hat \Omega}_{1}\over \bx-{1\over 2}}+{{\hat \Omega}_2\over \bx+{1\over 2}},\label{D_oper}
		\ee
		and here 
		\be
		{\hat \Omega}_1={\left(\p_{t}+(\Om^{R}_{+}+\Om^{L}_{+})\p_{\vp}+(\Om^{R}_{+}-\Om^{L}_{+})\p_{\psi}\right)^2 \over 4\kappa^2_{+}}, \quad {\hat \Omega}_2={\left(\p_{t}+(\Om^{R}_{-}+\Om^{L}_{-})\p_{\vp}+(\Om^{R}_{-}-\Om^{L}_{-})\p_{\psi}\right)^2 \over 4\kappa^2_{-}}.\label{Om_oper}
		\ee
		A systematic procedure of derivation of $SL(2,\mathbb{R})$ generators for near-zone approximation for Myers-Perry black hole was carried out in \cite{Charalam_2023} and, more recently, in~\cite{Gray}, and we now perform a similar derivation to the STU black role. Firstly, there are three simple Killing vectors for the considered metric, namely $\p_t$, $\p_{\vp}$ and $\p_{\psi}$, thus one of the generators can be chosen in the form \cite{Charalam_2023}:
		\be
		\hat{L}_{0}=-\left(\beta\p_{t}+\betap{\Om}_{+}\p_{+}+\betam{\Om}_{-}\p_{-}\right),
		\ee
		where $\beta$, $\betap$ and $\betam$ are constant parameters, $\Om_{+}$, $\Om_{-}$ correspond to our left and right angular velocities $\Om^{R}$ and $\Om^{L}$ respectively, and the derivative operators $\p_{\pm}\equiv\p_{\psi_{\pm}}={1\over 2}\left(\p_{\phi}\pm\p_{\psi}\right)$. The operators $\hat{L}_{\pm1}$ can be chosen in the form
		\be
		\hat{L}_{\pm 1}=G_{\pm}\p_{\bx}+K_{\pm}\p_{t}+H^{(+)}_{\pm}\Om_{+}\p_{+}+H^{(-)}_{\pm}\Om_{-}\p_{-}
		\ee
		where all the functions $G_{\pm}$, $K_{\pm}$, $H^{(+)}_{\pm}$ and $H^{(-)}_{\pm}$ in general depend on the four coordinates $t$, $\bx$, $\vp$ and $\psi$. Temporal dependence can be easily extracted from these functions if the following coordinate transformation is performed:
		\be 
		\tilde{\psi}_{\pm}=\psi_{\pm}-{\alpha_{\pm}\Om_{\pm}\over \beta}t, \quad \tilde{t}=t
		\ee
		where $\psi_{\pm}=\vp\pm\psi$. This transformation allows the  rewriting of the generator $\hat{L}_0$ in very simple form, namely,
		\be
		\hat{L}_0=-\beta{\partial\over{\partial\tilde{t}}}.
		\ee
		Using the constraint $[\hat{L}_{\pm 1},\hat{L}_{0}]=\pm\hat{L}_{\pm 1}$ and taking into account the fact that the Casimir (\ref{Casimir}) does not depend on angular variables explicitly, we find that the functions $X_{\pm}\equiv \left\{G_{\pm}, K_{\pm}, H^{+}_{\pm}, H^{-}_{\pm}\right\}$ can be chosen in the form
		\be
		X_{\pm}=e^{\pm\left({\tilde{t}\over{\beta}}+\tau_{+}\tilde{\psi}_{+}+\tau_{-}\tilde{\psi}_{-}\right)}{\cal X}_{\pm}(\bar{x}).
		\ee
		Thus, the generators $\hat{L}_{\pm 1}$ can be written as:
		\be
		\hat{L}_{\pm 1}=e^{\pm\left({\tilde{t}\over{\beta}}+\tau_{+}\tilde{\psi}_{+}+\tau_{-}\tilde{\psi}_{-}\right)}\left({\cal G}_{\pm}\partial_{\bar{x}}+{\cal K}_{\pm}\partial_{\tilde{t}}+\tilde{\mathcal{H}}^{(+)}_{\pm}\Om_{+}\tilde{\partial}_{+}+\tilde{\mathcal{H}}^{(-)}_{\pm}\Om_{-}\tilde{\partial}_{-}\right),
		\ee
		where $\tilde{\p}_{\pm}\equiv\p_{\tilde{\psi}_{\pm}}=\p_{\pm}$ and $\tilde{\mathcal{H}}^{(i)}_{\pm}=\mathcal{H}^{(i)}_{\pm}-{\alpha_{i}\over \beta}{\cal K}_{i}$, $i=\pm$. Casimir constraints then give rise to the following relations:
		\be 
		{\cal G}_{\pm}=\mp\sqrt{\bar{\Delta}}, \quad {\cal K}_{\pm}(\bar{x})={\cal K}(\bar{x}), \quad \tilde{\cal H}^{(i)}_{\pm}(\bar{x})=\tilde{\cal H}^{(i)}(\bar{x}).
		\ee
		
		The commutation relation $[\hat{L}_{+1},\hat{L}_{-1}]=2\hat{L}_{0}$ gives rise to the following equations for the functions ${\cal K}(\bar{x})$ and $\tilde{\cal H}^{(i)}(\bar{x})$:
		\begin{eqnarray}
			\sqrt{\bar{\Delta}}{\cal K}'(\bar{x})+{\cal K}(\bar{x})\left({{\cal K}(x)\over \beta}+\tau_{+}\tilde{\cal{H}}^{(+)}(\bar{x})+\tau_{-}\tilde{\cal{H}}^{(-)}(\bar{x})\right)=\beta,\label{equat_K}\\\sqrt{\Delta}\tilde{\cal H}^{(i)'}(\bar{x})+\tilde{\cal H}^{(i)}(\bar{x})\left({{\cal K}(x)\over \beta}+\tau_{+}\tilde{\cal{H}}^{(+)}(\bar{x})+\tau_{-}\tilde{\cal{H}}^{(-)}(\bar{x})\right)=0,\label{equat_hi}
		\end{eqnarray}
		where $'$ denotes the derivative with respect to $\bar{x}$. From the Casimir constraint (\ref{Casimir}) together with (\ref{D_oper}) we obtain that:
		\be
		\sqrt{\bar{\Delta}}\left({{\cal K}(x)\over \beta}+\tau_{+}\tilde{\cal{H}}^{(+)}(\bar{x})+\tau_{-}\tilde{\cal{H}}^{(-)}(\bar{x})\right)={\bar{\Delta}'\over 2}.\label{add_constr}
		\ee
		Combining the upper relation together with the equations (\ref{equat_K}), (\ref{equat_hi}) we can easily derive the explicit relations for the functions ${\cal K}(\bar{x})$ and $\tilde{\cal H}^{(i)}(\bar{x})$:
		\be
		{\cal K}(\bar{x})={{\beta\bar{x}+\delta}\over\sqrt{\bar{\Delta}}},\quad \tilde{\cal H}^{(i)}(\bar{x})={C^{(i)}\over \sqrt{\bar{\Delta}}},
		\ee
		where  $\delta$ and $C^{(i)}$ are integration constants. These integration constants  can be obtained if one uses explicit form of the operator (\ref{D_oper}) together with (\ref{Om_oper}). For computational convenience the operators can be rewritten in the form:
		\be
		\hat{\Omega}_{1}={1\over 4\kappa^2_{+}}\left(\partial_t+2\Omega^{R}_{+}\partial_{+}+2\Omega^{L}_{+}\partial_{-}\right)^2, \quad \hat{\Omega}_{2}={1\over 4\kappa^2_{-}}\left(\partial_t+2\Omega^{R}_{-}\partial_{+}+2\Omega^{L}_{-}\partial_{-}\right)^2.\label{Om_oper2}
		\ee
		The Casimir relation (\ref{Casimir}) now gives rise to the constraint:
		\be
		{\delta\over\beta}+\tau_{+}C^{(+)}+\tau_{-}C^{(-)}=0.
		\ee
		Using the expressions (\ref{Om_oper2}) we are able to derive explicit relations for the constants $\beta$, $\delta$ and $C^{(i)}$. The solution of the corresponding equations is not unique, and we thus obtain two distinct sets of operators whose commutation relations give rise to the required algebra. The first set of parameters is as follows:
		\begin{multline}
			\beta^{(1)}={1\over 2}\left({1\over \kappa_{+}}+{1\over \kappa_{-}}\right),\quad \delta^{(1)}={1\over 4}\left({1\over \kappa_{+}}-{1\over \kappa_{-}}\right), \quad \betap^{(1)}\Omega^{(1)}_{+}={2\Omega^{R}_{+}\over \kappa_{+}}, \\ \betam^{(1)}=0, \quad C^{(-)}_{(1)}={\Omega^{L}_{+}\over\kappa_{+}}, \quad C^{(+)}_{(1)}=-{\delta^{(1)}\over\beta^{(1)}}\betap^{(1)}\Omega^{(1)}_{+}={{\Omega^{R}_{+}(\kappa_{+}-\kappa_{-})}\over\kappa_{+}(\kappa_{+}+\kappa_{-})}.
		\end{multline}
		In particular, we note that the parameters $\beta^{(1)}$ and $\delta^{(1)}$ above are, except for numerical multiplicative factors, the very same constants $p$ and $q$ defined earlier. These constants are also the only way in which boost-dependence enters the generators, as all other parameters in the set above can be written in terms of ratios which, by~\eqref{eq.ok}, depend solely on the seed parameters.
		
		Similarly, we can use the derived parameters to write the operators $\hat{L}^{(1)}_{0}$ and $\hat{L}^{(1)}_{\pm 1}$ explicitly: 
		\begin{subequations}\label{Set1}
			\be
			\hat{L}^{(1)}_{0}=-\beta^{(1)}{\partial\over{\partial \tilde{t}_{1}}}=-\left(\beta^{(1)}\partial_t+\betap^{(1)}\Omega^{(1)}_{+}\partial_+\right)=-\left({(\kappa_{+}+\kappa_{-})\over 2\kappa_{+}\kappa_{-}}\partial_t+{2\Omega^{R}_{+}\over \kappa_{+}}\partial_{+}\right).
			\ee
			\begin{equation}
				\begin{split}
					&e^{\mp\left({t\over\beta^{(1)}}+\tau^{(1)}_{+}\tilde{\psi}^{(1)}_{+}+\tau^{(1)}_{-}\tilde{\psi}^{(1)}_{-}\right)}\hat{L}^{(1)}_{\pm 1}=\left(\mp\sqrt{\bar{\Delta}}\partial_{\bar{x}}+{1\over\sqrt{\bar{\Delta}}}\left(\left(\beta^{(1)}\bar{x}+\delta^{(1)}\right)\partial_{\tilde{t}_{1}}+C^{(+)}_{(1)}\partial_{+}+C^{(-)}_{(1)}\partial_{-}\right)\right)=\\ &\left(\mp\sqrt{\bar{\Delta}}\partial_{\bar{x}}+{1\over\sqrt{\bar{\Delta}}}\left({1\over 2\kappa_{+}\kappa_{-}}\left((\kappa_{+}+\kappa_{-})\bar{x}+{1\over 2}(\kappa_{-}-\kappa_{+})\right)\partial_t+{2\Omega^R_{+}\over \kappa_{+}}\bar{x}\partial_{+}+{\Omega^L_{+}\over \kappa_{+}}\partial_{-}\right)\right) \label{Lpm}.
				\end{split}
			\end{equation}
			
		\end{subequations}
		
		The second group of parameters is as follows:
		\begin{multline}
			\beta^{(2)}={1\over 2}\left({1\over \kappa_{+}}-{1\over \kappa_{-}}\right),\quad \delta^{(2)}={1\over 4}\left({1\over \kappa_{+}}+{1\over \kappa_{-}}\right), \quad \betam^{(2)}\Omega^{(2)}_{-}={2\Omega^{L}_{+}\over \kappa_{+}}, \\ \betap^{(2)}=0, \quad C^{(+)}_{(2)}={\Omega^{R}_{+}\over\kappa_{+}}, \quad C^{(+)}_{(2)}=-{\delta^{(2)}\over\beta^{(2)}}\betam^{(2)}\Omega^{(2)}_{-}={{\Omega^{L}_{+}(\kappa_{+}+\kappa_{-})}\over\kappa_{+}(\kappa_{+}-\kappa_{-})}.
		\end{multline}
		
		Now we find a structure very similar to that of the first set of parameters, but the roles of $p$ and $q$ are interchanged, with the former now proportional to $\delta^{(2)}$  and the latter to $\beta^{(2)}$. The generators for this set of parameters are
		\begin{subequations}\label{Set2}
			\begin{equation}
				\hat{L}^{(2)}_{0}=-\beta^{(2)}{\partial\over{\partial \tilde{t}_{2}}}=-\left(\beta^{(2)}\partial_t+\alpha^{(2)}_{-}\Omega^{(2)}_{-}\partial_-\right)=-\left({(\kappa_{-}-\kappa_{+})\over 2\kappa_{+}\kappa_{-}}\partial_t+{2\Omega^{L}_{+}\over \kappa_{+}}\partial_{-}\right)
			\end{equation}
			and
			\begin{multline}
				e^{\mp\left({t\over\beta^{(2)}}+\tau^{(2)}_{+}\tilde{\psi}^{(2)}_{+}+\tau^{(2)}_{-}\tilde{\psi}^{(2)}_{-}\right)}\hat{L}^{(2)}_{\pm 1}=\left(\mp\sqrt{\bar{\Delta}}\partial_{\bar{x}}+{1\over\sqrt{\bar{\Delta}}}\left(\left(\beta^{(2)}\bar{x}+\delta^{(2)}\right)\partial_{\tilde{t}_{2}}+C^{(+)}_{(2)}\partial_{+}+C^{(-)}_{(2)}\partial_{-}\right)\right)=\\ \left(\mp\sqrt{\bar{\Delta}}\partial_{\bar{x}}+{1\over\sqrt{\bar{\Delta}}}\left({1\over 2\kappa_{+}\kappa_{-}}\left((\kappa_{-}-\kappa_{+})\bar{x}+{1\over 2}(\kappa_{+}+\kappa_{-})\right)\partial_t+{\Omega^R_{+}\over \kappa_{+}}\partial_{+}+{2\Omega^L_{+}\over \kappa_{+}}\bar{x}\partial_{-}\right)\right).
			\end{multline}
		\end{subequations}

		We thus find two sets of generators for this algebra, with the corresponding Casimir operators satisfying eigenvalue equations of the form
		\begin{equation}
			\mathcal{\hat C}_{2}^{(j)}\Psi^{(j)}=\frac{\ell(\ell+2)}{4}\Psi^{(j)},
		\end{equation} 
		where $j=1,2$. A physical solution must also be consistent with the angular momentum eigenvalue equations
		\begin{equation}\label{angularmomenta}
			\hat J_{R/L}\Psi=-i\partial_{\pm}\Psi=m_{R/L}\Psi,
		\end{equation}
		where we have used the definitions $\p_{\pm}\equiv{1\over 2}\left(\p_{\phi}\pm\p_{\psi}\right)$ mentioned above to ensure that the eigenvalues at the right-side of~\eqref{angularmomenta} are indeed the same $m_R$ and $m_L$ that solve the eigenvalue of equation for the right and left angular momenta. Moreover, the action of $\hat L_{0}$ in the scalar wavefunctions gives, for each of the two sets of generators considered, the equations 
		\begin{equation}\label{weight1}
			\hat{L}^{(1)}_{0}\Psi^{(1)}=-i\left(-\beta^{(1)}\omega +2a_R\right)\Psi^{(1)}
		\end{equation}
		and 
		\begin{equation}\label{weight2}
			\hat{L}^{(2)}_{0}\Psi^{(2)}=-i\left(-\beta^{(2)}\omega +2a_{L}\right)\Psi^{(2)},
		\end{equation}
		where $a_R$ and $a_L$ were defined in Eq.~\eqref{eq.aRaL}. From the above eigenvalue equations, one can read the weights $-i\beta^{(1)}\omega +2a_R$ and $-i\beta^{(2)}\omega +2a_{L}$ relative to representations whose steps are generated by each set of operators. One very important result found in Ref.~\cite{CharalamPRL2021} for Kerr black holes, and recently generalized to the Myers-Perry solution~\cite{Charalam_2023} is the connection between highest-weight representations and the vanishing of static Love numbers, thus framing the absence of tidal deformations in black holes as selection rules enforced by the Love symmetry generators. Let us now verify that this connection still exists for the charged black holes considered in this work.
		
		In what follows, we shall for simplicity omit the numeric labels $1,2$ used above to distinguish between the two sets of generators, and note that the following calculations use the operators given by~\eqref{Set1}. Also, let us define the parameters $\hat{\ell}\equiv \frac{\ell}{2}$ which are often used in higher-dimensional spacetimes~\cite{PRD108, Charalam_2023} and turn out to be more convenient for the following discussion.  A highest-weight representation of weight $-\hat{\ell}$ is generated through the action of Ladder operators into the primary state $\Psi_{-\hat{\ell}}$, which must satisfy the conditions 
		\begin{subequations}\label{highestweight}
			\begin{align}
				\hat{L}_0 \Psi_{-\hat{\ell}} &=-\hat{\ell} \Psi_{-\hat{\ell}}  \label{L0cond} \\ 
				\hat{L}_1 \Psi_{-\hat{\ell}} &= 0. \label{Lpluscond}
			\end{align}
		\end{subequations}
		The highest-weight primary state can be found by solving the system of first-order equations implied by the above conditions. Integration of~\eqref{Lpluscond} with periodic angular dependence fixed through the requirement of consistency with ~\eqref{angularmomenta} gives

		\begin{equation}\label{FundamentalState}
			\Psi_{-\hat{\ell}} = N \, \left( \bar{x} - \frac{1}{2} \right)^{\frac{\hat{\ell}}{2} + B} \left( \bar{x} + \frac{1}{2} \right)^{\frac{\hat{\ell}}{2} - B} \exp\left( i m_{R} \psi_{+} + i m_{L} \psi_{-} + s t \right)
		\end{equation}
		where $N$ is a normalization constant, 
		
		\begin{equation}
			B = \frac{\hat{\ell} (\kappa_{-} - \kappa_{+})}{2(\kappa_{+} + \kappa_{-})} + i \left( \frac{a_{R} (\kappa_{+} - \kappa_{-})}{ (\kappa_{+} + \kappa_{-})} + a_{L} \right),
		\end{equation}
		and $s$ is fixed by fixing the weight of the presentation as $-\hat{\ell}$, i.e., by ensuring consistency with the eigenvalue equation~\eqref{L0cond}, which implies 
		\begin{equation}\label{sgen}
			s = \frac{2 \kappa_{+} \kappa_{-}}{\kappa_{+} + \kappa_{-}} \left( \hat{\ell} - i \frac{2 \Omega_{+}^{R}}{\kappa_{+}} m_{R} \right).
		\end{equation}
		
		By recognizing the relationship $s=-i\omega$, we can rewrite $B$ in the form
		\begin{equation}\label{Bgen}
			B = i \left( \frac{\omega (\kappa_{+} - \kappa_{-})}{4\kappa_{+}\kappa_{-}} + a_L \right).
		\end{equation}

		Thus, the primary state's radial dependence is made up of an oscillatory term of the form $N\left[(\bar x-1/2)/(\bar x+1/2)\right]^{i\operatorname{Im}(\emph{B})}$, which tends to unity at large $\bx$, multiplied by a factor of $(\bar x^2-1/4)^{\hat \ell/2}$, which grows as $(\bar x)^{\hat \ell}\sim r^{\ell}$. In particular,this oscillatory factor becomes $B = i a_L$ in the static case. Since this state is regular at the outer horizon, so are the descendant states derived from successive applications of $\hat L_{-1}$, as this operator is also regular. From the form of this operator, it is clear that, up to the complex prefactor inherited from $\Psi_{\hat\ell}$, descendant states behave as polynomials at infinity, with no decaying modes being created. In fact the associated asymptotic behavior is precisely of the quasi-polynomial implied by~\eqref{StaticLNvanishingcond}, so that this state is consistent with a static solution with vanishing Love numbers. Moreover, the large $\bx$ behavior of  descendant states is dominated by the action of the term proportional to $\sqrt{\bar\Delta}\partial_{\bx}\sim \bx\partial_{\bx}$ in~\eqref{Lpm}, which preserves the asymptotic properties in descendant states.  
		
		From the above discussion, it is clear that, if $\ell$ is an even integer, then the $\hat{\ell}$-th descendant of $\Psi_{-\hat{\ell}}$ displays the expected $r^{\ell}$ asymptotic behavior of the static $\Phi_{\ell}$ solution of the Klein-Gordon equation in this background. Let us now verify that, if $a_L=0$, this is indeed a solution of this equation, so that, from the aforementioned absence of decaying modes, follows the vanishing  of static Love numbers for states belonging to this highest-weight representation. First, we claim that the $k$-th descendant $(\hat L_{-1})^k\Psi_{-\hat{\ell}}$ is a weight $-\hat{\ell} + k$ state of this representation. Using the highest-weight conditions and the commutation relations~\eqref{SL2cr}, we find
		\begin{equation}
			\hat{L}_{0}\hat L_{-1}\Psi_{-\hat{\ell}}=\hat L_{-1} \hat L_{0}\Psi_{-\hat{\ell}} + [\hat L_{0},\hat L_{-1}]\Psi_{-\hat{\ell}}=\left(-\hat\ell +1\right)\hat L_{-1}\Psi_{-\hat{\ell}},
		\end{equation} 
		since $ [\hat L_{0},\hat L_{-1}]=L_{-1}$. If
		\begin{equation}\label{k_weight}
			\hat L_{0}(\hat L_{-1})^k\Psi_{-\hat{\ell}}=\left(-\hat\ell +k\right)(\hat L_{-1})^k\Psi_{-\hat{\ell}},
		\end{equation}
		then a similar derivation shows that $\hat L_{0}(\hat L_{-1})^{k+1}\Psi_{-\hat{\ell}}=(\hat L_{-1})^{k+1}\Psi_{-\hat{\ell}} + \hat{L}_{-1}\hat L_{0}\left((\hat L_{-1})^k\Psi_{-\hat{\ell}}\right)$, which equals $(-\hat\ell+k+1)(\hat L_{-1})^{k+1}\Psi_{-\hat{\ell}}$, thus proving~\eqref{k_weight} by finite induction. In particular, it follows that the $\hat{\ell}$ descendant is a weight zero state. However, the solution must be consistent with~\eqref{weight1}, hence, in the static case, 
		\begin{equation}\label{StaticVanishingCond}
			\hat L_{0}(\hat L_{-1})^{\hat{\ell}}\Psi_{-\hat{\ell}}=0\implies a_R=0,
		\end{equation}
		thus agreeing with our previous conclusion that null static Love numbers imply vanishing of either $a_L$ or $a_R$, with the vanishing of the former parameter being a consequence of an essentially identical derivation conducted for the second set of operators and using~\eqref{weight2}.
		
		We have now derived the necessary constraints on the parameters $\ell$, $a_{R}$ and $a_L$, but it remains to be shown that is indeed a solution of the static Klein-Gordon equation with the correct eigenvalue. To do this, we first note that $\Psi_{-\hat{\ell}}$ is an eigenstate of $\mathcal{\hat C}$ with eigenvalue $\hat{\ell}\left(\hat{\ell}+1\right)$. Indeed, 
		
		\begin{equation}
			\mathcal{\hat C}_2\Psi_{-\hat{\ell}}=\left(\hat{L}_{0}^2 -\frac{1}{2}[\hat{L}_1,\hat{L}_{-1}]\right)\Psi_{-\hat{\ell}}=\left(\hat{L}_{0}^2 -\hat{L}_{0}\right)\Psi_{-\hat{\ell}}=\hat \ell(\hat \ell+1)\Psi_{-\hat{\ell}}.
		\end{equation}
		
		Since the Casimir operator commutes with all generators by construction, it is immediately seen that the first descendant state and, by finite induction, all subsequent descendant states, are also eigenvalues of $\mathcal{\hat C}_2$ with eigenvalue $\hat \ell(\hat \ell+1)$. In particular we have, for $\hat \ell$-th descendant,
		
		\begin{equation}
			\mathcal{\hat C}_2\left \{\left((\hat L_{-1}^{\frac{\ell}{2}}\Psi_{-\frac{\ell}{2}}\right) \right \}= \frac{\ell( \ell+2)}{4}\left(\hat L_{-1}^{\frac{\ell}{2}}\Psi_{-\frac{\ell}{2}}\right),
		\end{equation}
		where we have returned to the original quantum number $\ell$. Thus, this state corresponds to a solution $\Phi_{\ell}$ of the static Klein-Gordon equation, with $\ell$ being an even integer and $a_R=0$. As remarked earlier, a largely identical derivation using the generators~\eqref{Set2} can be performed to prove the vanishing of static Love numbers for a state in the  corresponding highest-weight representation of the associated algebra, with the difference that~\eqref{weight2} now implies $a_L=0$, thus covering all conditions that lead to vanishing Love numbers, as discussed in Sec.~\ref{sec:solutions}. 
		
		The above derivation has emphasized the static case, as this is arguably the most important one in Love number theory. However, the fundamental state is also defined when $\omega\neq 0$ (and thus $s\neq 0$), so we can generalize this argument to the dynamic case. In fact, we have derived the fundamental state, and in particular the $\omega$-dependent parameter $B$, without assuming $\omega=0$, so this state is valid in general. Moreover, the proof above relies very little on the specific form of the operators, as these results mostly follow from algebraic properties. Thus, we can set up a highest-weight representation by using $s$ given by~\eqref{sgen} (obtained from the weight condition for the fundamental state) and~\eqref{Bgen} to show that the $\hat{\ell}$-th descendant state is an appropriate solution of the dynamic wave equation in the near-zone regime, as implied by the Casimir equation  through the same algebraic derivation as before. The crucial difference comes in the weight zero constraint for this solution, which is now generalized, in the case of the first set of operators, to

		\begin{equation}
			-\beta^{(1)}\omega +2a_R =0\implies	\frac{\omega}{4} p = a_{R},
		\end{equation}
		where we have inserted the definitions of $p$ and $a_R$ into the first equation to derive the second one. For the remaining set of operators, we have
		\begin{equation}
			-\beta^{(2)}\omega +2a_L =0\implies	\frac{\omega}{4} q = a_{L}.
		\end{equation}	
		
		The above equations are \emph{exactly} the conditions~\eqref{DynamicalLNVanishingGen} we have derived by imposing absence of RG flow. Not only does this result serve as a check for our previous calculations (as it has been derived through a largely independent method), it is also relevant for revealing the connection between Love symmetry and Love number vanishing in the dynamical scenario, putting~\eqref{DynamicalLNVanishingGen} and~\eqref{StaticLNvanishingcond} in equal footing. In this language, the near-zone conditions~\eqref{DynamicalVanishingCond} are interesting in that they lead to solutions belonging to a highest-weight representation for \emph{all} $\omega$.

		In this section, we have shown that the Love symmetry structure of the seed Myers-Perry solution is consistently preserved by the generation technique that leads to STU black holes, with appropriately generalized boost-dependent  generators for the algebra of $SL(2,\mathbb{R})$. Vanishing conservative responses can be explained by the fact that the related solutions belong to a highest-weight representation produced by the Love-Symmetry generators, thus generalizing the findings of Ref.~\cite{Charalam_2023} for Myers-Perry black holes. Moreover, the above results provide a physical distinction between the two sets of generators we have found, as each of them is associated with the vanishing of one of the left/right-handed angular momentum parameters. Interestingly, we have found these results by a slightly different method than the one used in Ref.~\cite{Charalam_2023}, where these two conditions correspond to distinct near-zone splits. Here, we have used only one split, but the same conditions ensued from the two sets of operators derived by imposing consistency between the Casimir and near-zone equations.

		\section{Conclusions}
		
		In this work, we have thoroughly investigated the tidal responses of STU black holes parameterized by mass, three charges and two independent angular momenta. We found the exact static solutions of the massless Klein-Gordon equation, which are found to be equivalent to the uncharged case due to~\eqref{eq.ok}. Time-dependent perturbations have also been investigated with use of the near-zone approximation, as is usual in black hole tidal theory. Through use of asymptotic analysis of the solutions, we have derived the full response coefficients and, using analytic continuation, derived the canonical Love numbers and dissipative coefficients of the black hole. We have also investigated the conditions that imply vanishing Love numbers. In the static case, these conditions are the same as those found in Ref.~\cite{Charalam_2023} for the Myers-Perry solution, a result that we have shown to remain valid even when charges are introduced. This situation is analogous to what is found in the four-dimensional theory, where static results have similarly been shown to depend only on the parameters of the seed Kerr solution.

		The dynamical results were found to be boost-dependent, showing that the U(1) charges affect the dynamical tidal responses. Remarkably, we found that these boosts only appear in the solutions through the variables $q$ and $p$ defined in~\eqref{Def:p_and_q}. Up to a $(2\pi)^{-1}$ factor, these parameters are in fact just the left and right inverse temperatures~\eqref{eq:betal} and~\eqref{eq:betar} respectively. In the dynamical case, there are some interesting  nontrivial situations leading to vanishing Love numbers, whether in approximate or exact sense. In particular, we found that, if one of these inverse temperatures vanishes, as happens in particular when the BPS limit is approached, the conditions that imply the vanishing of static Love numbers are automatically generalized to arbitrary frequencies.  Specifically, we have verified that, when the BPS bound is approached, the conditions $a_{R/L}=0$, $\frac{\ell}{2}\in\mathbb{N}$ lead to Love number vanishing for all $\omega$ in the near-zone approximation. It is remarkable that the seemingly unrelated requirement of (near) vanishing Love numbers seems to suggest very naturally the same conditions that lead to (near) BPS saturation. This is an interesting feature that has no analogue in the neutral black holes that have been studied the most, as the Bogomol'nyi bound becomes trivial when all U(1) charges are set to zero. We also note that conditions~\eqref{DynamicalVanishingCond} lead to special scenarios with a clear thermodynamic interpretation, which is is an interesting feature that merits further investigation. 

		We have also developed a ladder formalism which can be used to relate the static solutions to each other through raising or lowering the value of $\ell$. These ladder operators lead to a Hamiltonian formulation of the wave equation and an associated algebra leading to conserved currents. These currents and Ladder operators can be derived directly from the Hamiltonian and the intertwining relations~\eqref{LadderAlgebra}, and may thus be useful tools that could be used to match the behavior of the wavefunction in different regions of space-time, even without prior knowledge of the exact solutions. This may be an important asset in other systems where closed-form solutions may not be achievable, but to which the Ladder formalism may be applied.  We have also verified that the zeroes of these currents coincide with those of the intrinsic dissipative response of the black hole, suggesting a connection between the two quantities. The vanishing of static Love numbers under conditions~\eqref{StaticLNvanishingcond} can be easily explained in this framework. Indeed we see that, under these conditions, the fact that regular solutions of the wave equation can be raised from a constant solution (which thus lacks a decaying branch) suffices to demonstrate the emergence of a polynomial behavior for these states. An asymptotic analysis of all other solutions using the same formalism shows that a polynomial form is not achieved when these conditions are violated. It is also notable that a Kerr-like ladder structure emerges in the cases where Love number vanishing is achieved.  These results can also be generalized to the dynamical case within the near-zone approximation. Although we did not perform this generalization explicitly in this work, the near-zone equation and solutions are seen to have the exact same form as in the static case. For this reason, a near-zone generalization of this ladder structure can be achieved in a straightforward manner, by shifting $a$, $b$ and $c$ to their dynamical values according to~\eqref{shifts}, and noting that currents and ladder operators can all be written in terms of these parameters, as seen form relations~\eqref{eq:gcr} and~\eqref{Rel_lowering}. Although this kind of ladder formalism has been developed for various black holes in four dimensions, \textcolor{black}{we do not know of any examples in the five-dimensional setting, so these results appear to be novel even for simpler particular cases such as the Schwarzschild-Tangherlini and Myers-Perry solutions.} Moreover, we note that conditions~\eqref{StaticLNvanishingcond} allow for an identification of the even $\ell$ states with analogue four-dimensional solutions of the Klein-Gordon equation in the STU background. In this sense, some of our results, notably in the static case, also apply to the $4D$ black holes investigated in Ref.~\cite{CveticPRD105}, and are in agreement with the results of that reference.

		Finally, we show that the Love symmetry structure found in the seed Myers-Perry solutions is preserved by the solution-generated technique, so that this symmetry persists regardless of charge. We derive the generators of the associated algebra of $SL(2,\mathbb{R})$ relevant to the STU solution and the Casimir corresponding to the near-zone truncation of the massless Klein-Gordon equation, and show that the vanishing of Love numbers in the special cases discussed can be explained by Love symmetry.  We investigate the conditions leading to vanishing Love numbers in both the static and dynamical settings, and show that these constraints amount to requiring that the solution be a weight zero state of the appropriate highest-weight representation.   
		
		Several extensions of the present work are possible. Here, we have worked with perturbations that solve the massless Klein-Gordon equation in the static and near-zone regimes, and thus dealt with \emph{scalar} Love numbers. This is a common practice, as it is well-known that the main features of tidal perturbations are largely spin-independent at the qualitative level, but it would still be desirable to generalize this result for different spins using a Teukolsky-type wave equation. As of this time, a master equation of this kind is not available for general STU black holes, but should future advancements enable its discovery, it would be worthwhile to extend our results to this scenario. Another potential generalization lies in the extension of this analysis to six or more dimensions. STU black hole solutions have been derived in higher-dimensional settings~\cite{NearBPSD} and present similar features, so an extension should be straightforward, although mathematically more challenging. We could also attempt to re-derive the conclusions of the present paper with use of the EFT formalism. This is a powerful framework, and perhaps more widely applicable than the analytic continuation method used here, since the latter would be difficult to use without closed-form solutions. The results thus obtained should be largely the same and, since the near-zone equation we are using has a relatively simple hypergeometric form, we found it preferable to use the arguably simpler framework of differential equation theory. That said, an EFT investigation may still bring new interpretations to some of our results, or perhaps bring light to features we have missed. Finally, another worthwhile path of future investigation may be encountered in the study of quasinormal modes and their properties.
		
		\section*{Acknowledgments}{One of the authors (MC) is partially supported by the Slovenian Research Agency (ARRS No. P1-0306), Fay R. and Eugene L. Langberg Endowed Chair funds, by DOE Award (HEP) DE-SC0013528, by the Simons Foundation Collaboration grant $\# 724069$ and by a University Research Foundation Grant at the University of Pennsylvania. MAL is funded by the Brazilian agency Conselho Nacional de Desenvolvimento Cient\'ifico e Tecnol\'ogico (CNPq), grant $\# 151204/2024-1$. M.C. would like to thank Gary Gibbons, Chris Pope and Bernard Whiting for collaboration on related topics.}

	\end{document}